\documentclass[journal abbreviation, nolinenumbers]{copernicus}

\usepackage{listings}             
\usepackage{color}
\definecolor{deepblue}{rgb}{0,0,0.5}
\definecolor{deepred}{rgb}{0.6,0,0}
\definecolor{deepgreen}{rgb}{0,0.5,0}
\definecolor{mygrey}{rgb}{0.5,0.5,0.5}

\defcitealias{itu_m2101_0}{ITU-R~Rec.~M.2101}
\defcitealias{itu_p452_14}{ITU-R~Rec.~P.452 v14}
\defcitealias{itu_p452_16}{ITU-R~Rec.~P.452}
\defcitealias{itu_p676_10}{ITU-R~Rec.~P.676}
\defcitealias{itu_p835_5}{ITU-R~Rec.~P.835}
\defcitealias{itu_p2108_0}{ITU-R~Rec.~P.2108}
\defcitealias{itu_ra769_2}{ITU-R~Rec.~RA.769}

\defcitealias{itu_tg5_1_doc36_att2}{ITU-R~Doc.~TG\,5/1-36 (Attachment 2)}
\defcitealias{itu_tg5_1_doc92_annex01}{ITU-R~Doc.~TG\,5/1-92 (Annex 1)}

\defcitealias{3gpp_tr_38.901}{3GPP~TR\,38.901}

\defcitealias{ecc_report_247}{ECC~Report~247}
\defcitealias{ecc_report_252}{ECC~Report~252}

\begin{document}

\title{Spectrum management and compatibility studies with Python}


\Author[1]{Benjamin}{Winkel}
\Author[1]{Axel}{Jessner}

\affil[1]{Max-Planck-Institut f\"{u}r Radioastronomie, Auf dem H\"{u}gel 69, 53121 Bonn, Germany}


\runningtitle{Spectrum management and compatibility studies with Python}

\runningauthor{B. Winkel \& A. Jessner}

\correspondence{Benjamin Winkel (bwinkel@mpifr.de)}

\received{}
\pubdiscuss{} 
\revised{}
\accepted{}
\published{}


\firstpage{1}

\maketitle
\nolinenumbers

\begin{abstract}
We developed the \texttt{pycraf} Python package, which provides functions and procedures for various tasks related to spectrum-management compatibility studies. This includes an implementation of \citetalias{itu_p452_16}, which allows to calculate the path attenuation arising from the distance and terrain properties between an interferer and the victim service. A typical example would be the calculation of interference levels at a radio telescope produced from a radio broadcasting tower. Furthermore, \texttt{pycraf} provides functionality to calculate atmospheric attenuation as proposed in \citetalias{itu_p676_10}.

Using the rich ecosystem of scientific Python libraries and our \texttt{pycraf} package, we performed a large number of compatibility studies. Here, we will highlight a recent case study, where we analysed the potential harm that the next-generation cell-phone standard 5G could bring to observations at a radio observatory. For this we implemented a Monte-Carlo simulation to deal with the quasi-statistical spatial distribution of base stations and user devices around the radio astronomy station.

\end{abstract}

\copyrightstatement{TODO}

\introduction  
The electromagnetic spectrum, especially at radio frequencies, is a limited resource, which has tremendous economic and scientific value for a large variety of services. Applications involve mobile communication networks, radio and TV broadcasting, RADAR, and safety distress signals. In contrast, natural sciences such as radio astronomy, Earth sensing (which is important for climatology), or weather forecasting rely heavily on interference-free spectral bands to conduct measurements with their high-sensitivity receivers. Especially, the radio-astronomy service (RAS) developed receivers, often cryogenically cooled, that allow us to detect even the faintest signals from outer space. Today it is possible to observe Micro-Jansky sources, with a Jansky (Jy) being a unit of the spectral power flux density, which was introduced to honour Karl. G. Jansky the ``founder'' of the radio-astronomy field. It is $1~\mathrm{Jy}=10^{-26}~\mathrm{W}\,\mathrm{m}^{-2}\,\mathrm{Hz}^{-1}$.

With the increased utilization of the radio spectrum, it became essential to regulate its usage. Allocations of frequency bands are nowadays made by the International Telecommunication Union Radiocommunication Sector (ITU-R) in unanimous decisions by delegates of the administrations of its member states. Every two to four years, ITU-R organizes the World Radiocommunication Conference (WRC) where the so-called Radio Regulations are updated, which contain rules and procedures, and spectrum allocation tables. It is clear that such decisions and allocations have to be made in accordance to the technical needs and constraints of the concerned services. Therefore, in many regional and international organizations and institutions compatibility studies are performed, which analyse if, and possibly under which restrictions, a service can be granted a new allocation.

In the following, we present a new Python library, called \texttt{pycraf}, that can be used to carry out many of the recurring tasks, which occur in compatibility studies. This includes the estimation of path propagation losses and atmospheric dampening effects, the determination of satellite positions (given orbital parameters), or calculating antenna gain patterns. After a short introduction into compatibility studies in Section~\ref{sec:compatstudy}, the more important \texttt{pycraf} features are briefly introduced in Section~\ref{sec:pycraf} along with basic usage examples. A real-world application is discussed in Section~\ref{sec:5g}, where the potential impact of the next-generation 5G mobile communication networks, which may acquire a new allocation at 24~GHz at the upcoming WRC-19, on the RAS is studied. Conclusions are presented in Section~\ref{sec:conclusions}.

\section{Spectrum management and compatibility studies}\label{sec:compatstudy}

Many compatibility studies follow a very simple recipe. The first step is usually to calculate the radiated power (or the power flux density), $P_\mathrm{tx}$, at the transmitter (Tx). Then one has to determine the appropriate interference threshold for the detected power, $P_\mathrm{rx}$, at the victim receiver (Rx). The ratio $P_\mathrm{tx}/P_\mathrm{rx}$ constitutes the minimum shielding between interfering Tx and victim Rx, also called ``minimum coupling loss'' (MCL).
In the final step the product of antenna gains and the path propagation loss (or path attenuation), i.e., the loss of power on the path to the receiver (Rx) is determined and compared to the MCL. That way one may find the minimum distance between a single interferer and victim receiver, which is often called separation distance. Since the path propagation loss also depends on the terrain height profile between Tx and Rx, the necessary separation distance will vary for different bearings, leading to non-circular exclusion zones around a victim receiver. In the case of a widespread deployment of many possible interferers one may need to calculate the expected total (``aggregated'') interference levels from all interferers within a certain range around the victim receiver. In most cases this will require a statistical simulation of possible deployments in a given terrain. However, simplified numerical convolutions of path loss laws with the deployment distribution functions are often cheaper to compute and can also yield good estimates of the size of zones where interferers ought to be excluded.

\subsection{Available software}
The various tasks that have to be solved in typical compatibility studies are often similar. Nevertheless, major effort is necessary to develop a single tool, which could be used in all sorts of different scenarios.

One such tool is \textsc{Seamcat} \citepalias{ecc_report_252},
which was developed by the European Communications Committee (ECC)
to be used by spectrum managers in  CEPT\footnote{The European Conference of Postal and Telecommunications Administrations (CEPT)
is an organization where policy makers and regulators from 48 countries across Europe collaborate to harmonize
telecommunication, radio spectrum and postal regulations. The European Communications Office (ECO) is the Secretariat of the CEPT.}
countries. It lets the user choose from a large amount of system parameters, deployment scenarios, and path propagation models -- all with appropriate distribution functions --
and applies Monte-Carlo sampling to calculate the desired output distributions, e.g., of the power levels detected at a receiver. Unfortunately, as of today, \textsc{Seamcat} can only perform so-called generic studies, i.e., the terrain profile between a transmitter and receiver is either neglected or approximated in a very coarse manner. For RAS stations in mountainous terrain this may lead to unrealistic  results as the diffraction at hill or mountain tops can significantly increase the path attenuation between victim and interferer(s).

PathProfile is a publicly available\footnote{\url{http://www.mike-willis.com/software.html}}
software tool for the prediction of local field strengths developed by Mike Willis. It features a GUI and uses terrain data for the calculation of path loss. As such, it may serve as the core of compatibility calculations, but will require additional specific interfacing for multi-interferer scenarios or aggregation studies.

A variety of MATLAB-based solutions to various spectrum management problems exists and is used by regulatory administrations, industry, and also by radio astronomers in spectrum management. Some of them have found approval by the ECC \citepalias[e.g.][]{ecc_report_247},
but many are not in the public domain and less well known and are used mainly by  individuals.

\section{The \texttt{pycraf} package}

While there are some existing software tools to aid with compatibility studies, they are either not suited for all relevant tasks (PathProfile), not easy to use from within a programming language (\textsc{Seamcat}), or are based on proprietary software (MATLAB-based scripts). Therefore, we decided to provide a number of functions and procedures, which were developed in the framework of several radio compatibility studies involving the Effelsberg observatory. The observatory is located in the Eifel mountains in Western Germany and operates a 100-m radio telescope. The code comes in the form of a Python library (a so-called package) that can easily be installed and used even by relatively inexperienced Python users. The package is named \texttt{pycraf}, open-source licensed (GPL-v3), and available on the Python package distribution server PyPI (\textit{Python Package Index}\footnote{\url{https://pypi.python.org/pypi/pycraf}}.
The source code is furthermore hosted on the GitHub platform\footnote{\url{https://github.com/bwinkel/pycraf}}, along with detailed documentation\footnote{\url{https://bwinkel.github.io/pycraf/}}, a bug tracker, and installation instructions.

We aim to make \texttt{pycraf} a community-driven project. Contributions to the source code or documentation are very welcome, but we are also happy to receive bug reports, feature requests and other proposals that would further improve the package.

The \texttt{pycraf} package makes use of the AstroPy package template providing a sophisticated build environment, which is based on \texttt{setuptools}, \texttt{pytest}, and \texttt{sphinx}. It allows the use of automated testing, makes incorporation of continuous integration services easy, and enables automated documentation generation. Because \texttt{pycraf} is available on the PyPI distribution server, one can use the Python tool \texttt{pip} to install the package into the local Python environment.

After successful installation, one can load and test the library with the following two statements:

\begin{lstlisting}[numbers=none,frame=,caption=]
import pycraf
pycraf.test(remote_data='any')
\end{lstlisting}
With this \texttt{remote\_data} option, \texttt{pycraf} will attempt to download terrain-height data from the NASA Shuttle Radar Topography Mission \citep[SRTM][]{farr07} for the calculation of path propagation loss in real terrain\footnote{See \url{https://bwinkel.github.io/pycraf/pathprof/working_with_srtm.html} for further details.}.

In the following we will present an overview of the features included in \texttt{pycraf} and provide some basic examples for its use. A discussion of all aspects of \texttt{pycraf} and introduction to the various third-party libraries used in \texttt{pycraf} is beyond the scope of this paper and we refer to the online manual where a lot more information and a full API reference for each sub-package can be found. We also note that \texttt{pycraf} makes extensive use of the Astropy (physical) \texttt{units} module, which is well documented on the Astropy website\footnote{\url{http://docs.astropy.org/en/stable/units/}}.

\subsection{Conversions module}\label{sec:pycraf}

Most of the fundamental operations, e.g., converting the radiated total power into an equivalent power flux density at a given distance and assuming a certain antenna gain of the Tx, or calculating the total received power at an Rx from the power flux density, are covered by the \texttt{conversions} sub-package. Furthermore, several Decibel units are defined, e.g.:

\begin{lstlisting}[frame=,caption=Conversions module.]
import astropy.units as u
import pycraf.conversions as cnv

power = 1 * u.W

print('{:.2f}'.format(power))
# 1.00 W

print('{:.2f}'.format(power.to(cnv.dB_W)))
# 0.00 dB(W)

print('{:.2f}'.format(power.to(cnv.dBm)))
# 30.00 dB(mW)
\end{lstlisting}
To calculate the power flux density (in vacuum) produced by a 1-W Tx at a distance of e.g. 1~km, assuming a forward antenna gain of 25~dBi and free-space propagation, one can use the function \texttt{powerflux\_from\_ptx}:

\begin{lstlisting}[firstnumber=last,frame=]
p_tx = 1 * u.W
gain_tx = 25 * cnv.dBi
distance = 1 * u.km

pfd = cnv.powerflux_from_ptx(
    p_tx, distance, gain_tx
    )

print('{:.2e}'.format(pfd))
# 2.52e-05 W / m2

print('{:.2f}'.format(
    pfd.to(cnv.dB_W_m2)
    ))
# -45.99 dB(W / m2)
\end{lstlisting}
Likewise, if such power flux density is received by an isotropic antenna, one can calculate the total received power (which depends on frequency):
\begin{lstlisting}[firstnumber=last,frame=]
gain_rx = 0 * cnv.dBi
frequency = 1 * u.GHz

p_rx = cnv.prx_from_powerflux(
    pfd, frequency, gain_rx
    )

print('{:.2e}'.format(p_rx))
# 1.80e-07 W

print('{:.2f}'.format(p_rx.to(cnv.dB_W)))
# -67.45 dB(W)
\end{lstlisting}

The difference between Tx and Rx power for isotropic antennas is also called the free-space loss and can be calculated using Friis' transmission equation:
\begin{lstlisting}[firstnumber=last,frame=]
fspl = cnv.free_space_loss(
    distance, frequency
    )

print('{:.2f}'.format(fspl))
# -92.45 dB
\end{lstlisting}
This result is consistent with the above approach, if the additional Tx antenna gain of 25~dBi is considered.

The \texttt{conversions} sub-package offers a lot more and we refer to the online documentation for a detailed overview of the functionality.

\subsection{Path attenuation}
The largest part of the code base is currently dedicated to the implementation of the path propagation loss algorithm described in \citetalias{itu_p452_16}.
As discussed in Section~\ref{sec:compatstudy}, the path attenuation calculation is one of the three fundamental steps in any compatibility study. Furthermore, it is the most critical aspect, as the losses can easily exceed 100~dB and typically uncertainties are large.
The method in P.452 includes various effects that influence the path loss:
\begin{enumerate}
\item Line-of-sight (free-space) loss including correction terms for multipath and focussing effects,\\[-4ex]
\item Diffraction (at terrain features),\\[-4ex]
\item Tropospheric scatter, and\\[-4ex]
\item Anomalous propagation (ducting, reflection from elevated atmospheric layers).
\end{enumerate}
Furthermore, a simple correction is provided to include additional losses caused by local obstacles (``clutter'') at the end points of the propagation path. The details of the individual mechanisms are beyond the scope of this paper and we refer to the P.452 and references therein for further information.

We note, however, that the diffraction calculation was revised in version 15 of the P.452. Previously, it used the 3-edge Deygout method, while the new algorithm is based on the so-called delta-Bullington method. Because this change can significantly affect the results of the diffraction loss in mountainous terrain \citep[see][and references therein]{wilson08}, \texttt{pycraf} offers both methods, from the older version 14 and the current version 16, such that one can compare older study results with the predictions of the new method.

While most of \texttt{pycraf} is pure Python code, for performance reasons the \texttt{pathprof} sub-package contains a large amount of Cython\footnote{Cython is an extension to the Python language that allows the compilation of code into so-called C~extensions, which can then be imported into Python and often provide a significant speed gain compared to Python-only implementations. The details of this are beyond the scope of this paper. More information can be found on the Cython webpage: \url{http://cython.org/}.} code \citep{Cython}. Even then, the P.452 algorithm is computationally demanding and if one wants to create a large map of path attenuations around a transmitter, the calculation would be relatively slow. Allowing for an insignificant loss of numerical accuracy, we developed a fast alternative for this use case, which is presented in Section~\ref{subsubsec:fastattenmap}.

\begin{figure}[t]
\includegraphics[width=8.3cm]{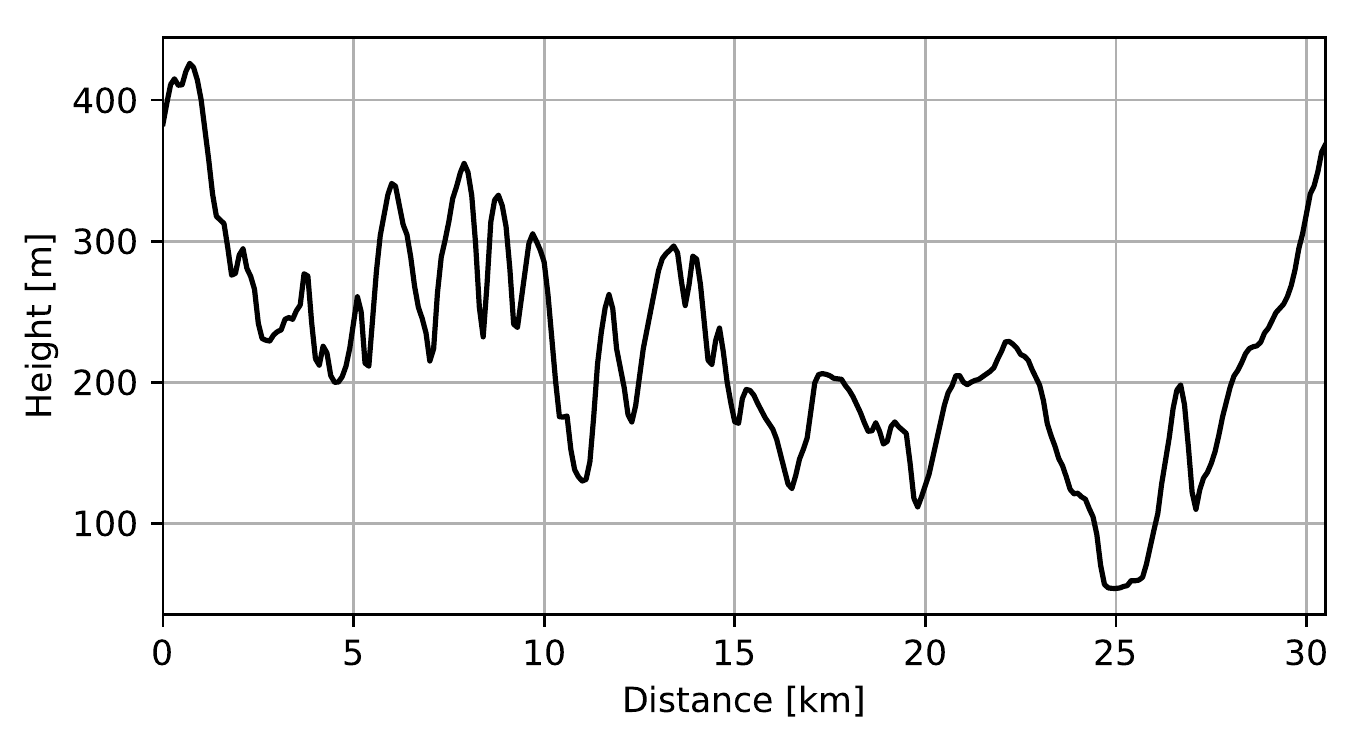}
\caption{Example height profile as obtained with the program in Listing 2. Based on NASA SRTM data \citep{farr07}.}
\label{fig:heightprof}
\end{figure}

Before the path propagation loss can be calculated, a terrain height profile has to be constructed from a suitable data source. The \texttt{pycraf} package has built-in access to SRTM data servers, which allows to obtain so-called tiles ($1^\circ\times1^\circ$) on demand:

\begin{lstlisting}[frame=,caption=Path profile module.]
import astropy.units as u
import pycraf.conversions as cnv
import pycraf.pathprof as pp

# allow download of missing SRTM data:
pp.SrtmConf.set(download='missing')

lon_t, lat_t = (6.9, 50.5) * u.deg
lon_r, lat_r = (7.3, 50.6) * u.deg
step = 100 * u.m

(
    lons, lats, dist, dists, heights,
    bearing, bbearing, bbearings
    ) = pp.srtm_height_profile(
        lon_t, lat_t, lon_r, lat_r, step
        )
\end{lstlisting}
Of course, it is also possible to provide SRTM tiles in a local directory to avoid excessive network traffic. The returned variables contain the geographical longitudes and latitudes of the path, the distances and back-bearings of each point on the path with respect to the transmitter \citep[which are inferred via Vincenty's formulae for the Earth ellipsoid;][]{vincenty_1975}, as well as the heights (above the WGS84 mean sea level). Here, the code to plot the data is omitted, but Fig.~\ref{fig:heightprof} displays the result.

It is noted, however, that the relevant \texttt{pycraf} functions can query SRTM data automatically; users do not necessarily need to provide the height profile themselves. Calculating the propagation loss between two points is a two-step process. First, an auxiliary object, \texttt{PathProp}, is constructed that contains the geometrical, environmental, and terrain parameters necessary for the further calculations:
\begin{lstlisting}[firstnumber=last,frame=]
frequency = 1 * u.GHz
omega = 0 * u.percent
temperature = 290 * u.K
pressure = 1013 * u.hPa
time_percent = 2 * u.percent
h_tg, h_rg = (5, 50) * u.m

zone_t = pp.CLUTTER.URBAN
zone_r = pp.CLUTTER.SUBURBAN

pprop = pp.PathProp(
    frequency,
    temperature, pressure,
    lon_t, lat_t,
    lon_r, lat_r,
    h_tg, h_rg,
    step,
    time_percent,
    zone_t=zone_t, zone_r=zone_r
    )
\end{lstlisting}
Here, \texttt{omega}, is the fraction of the path that is over sea, \texttt{time\_percent} is the percentage of time for which the propagation losses will be larger than the returned values (see P.452 for details), and $h_\mathrm{tg,rg}$ are the antenna heights of the Tx and Rx over ground, respectively. Furthermore, the clutter zone type can be specified for both end points.

\begin{figure}[t]
\includegraphics[width=8.3cm]{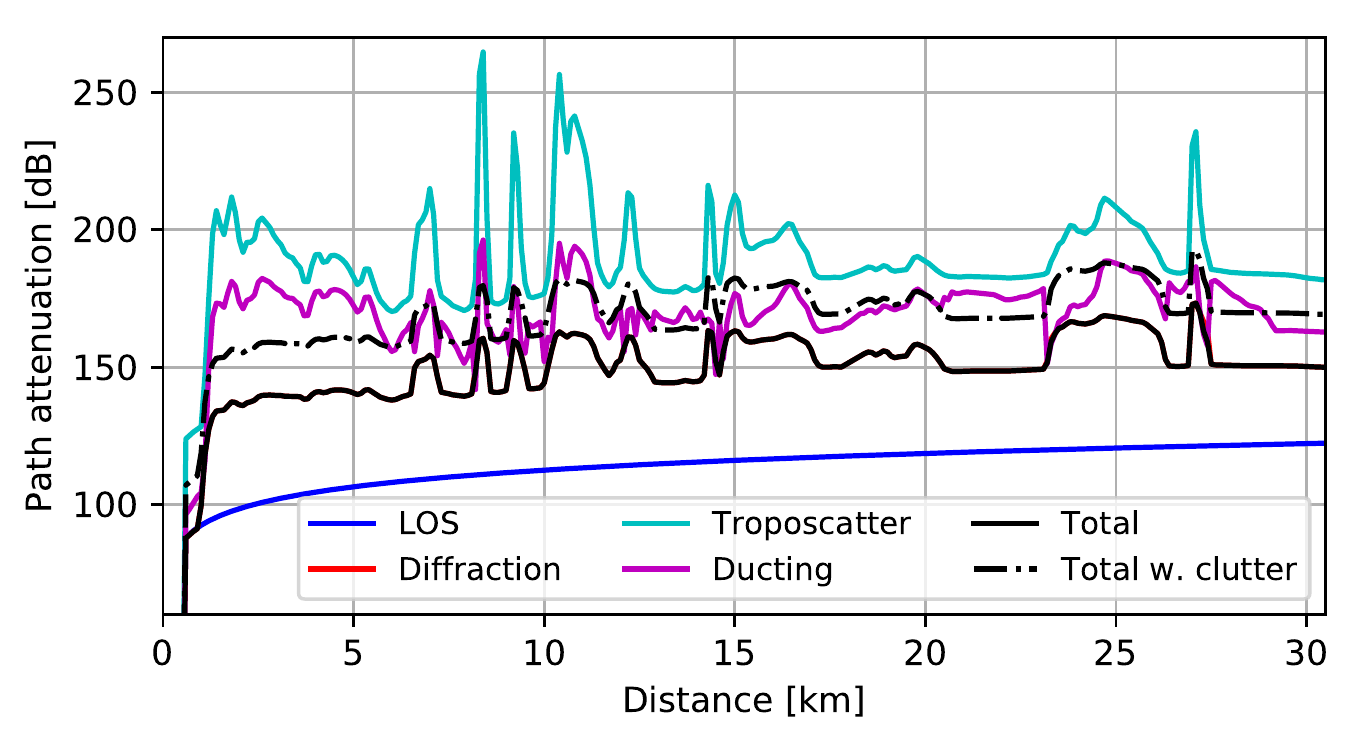}
\caption{Path propagation loss along the terrain height profile displayed in Fig.~\ref{fig:heightprof}. Based on NASA SRTM data \citep{farr07}.}
\label{fig:pathatten}
\end{figure}

The next step consists of  feeding the \texttt{PathProp} instance into a function \texttt{loss\_complete}
\begin{lstlisting}[firstnumber=last,frame=]
# account for additional antenna gains
G_t, G_r = (20, 15) * cnv.dBi

tot_loss = pp.loss_complete(
    pprop, G_t, G_r
    )
for l in tot_loss:
    print('{:.2f}'.format(l))
# 122.34 dB
# 149.94 dB
# 181.99 dB
# 162.70 dB
# 149.94 dB
# 169.15 dB
# 134.15 dB
\end{lstlisting}
The returned Python \texttt{tuple} contains
\begin{enumerate}
\item $L_\mathrm{bfsg}$, the free-space loss,\\[-4ex]
\item $L_\mathrm{bd}$, the basic transmission loss caused by diffraction,\\[-4ex]
\item $L_\mathrm{bs}$, the tropospheric scatter loss,\\[-4ex]
\item $L_\mathrm{ba}$, the ducting/layer reflection loss,\\[-4ex]
\item $L_\mathrm{b}$, the complete path propagation loss,\\[-4ex]
\item $L_\mathrm{b,corr}$, as $L_\mathrm{b}$ but with clutter correction, and\\[-4ex]
\item $L$, as $L_\mathrm{b}$ but with clutter and gain correction.
\end{enumerate}
An important property of the total propagation loss ($L_\mathrm{b}$ and $L$) is that it is not simply the product (or sum, if logarithmic representation is used) of the components. Again, we refer to P.452 for more details. To give the reader an impression on how the various quantities behave as a function of distance from the transmitter, we have calculated the components for each point in the terrain height profile (Fig.~\ref{fig:heightprof}) and display them in Fig.~\ref{fig:pathatten}. In contrast to the two-point calculation above, the antenna gains were neglected, as the effective gain towards each point along the path differs somewhat due to the different elevation angles of the path at the two stations. That however requires accounting for the respective antenna patterns, which is supported by \texttt{pycraf}, but beyond the scope of this simple example.




\subsubsection{A fast map-making algorithm}\label{subsubsec:fastattenmap}
With the above approach it is straightforward to produce a path loss map, e.g., around a given transmitter, by simply calculating the propagation loss between each pixel in the map and the Tx location. As the P.452 algorithm is relatively complex, this would however require a substantial amount of computing time. In addition, the construction of the geodesics (the shortest paths between two points on the Earth ellipsoid) and SRTM height profile extraction/interpolation are also computationally demanding. Thus, two additional techniques were introduced to provide a significant speed-up. First, the map computation was moved into the Cython part of the code (avoiding the use of slower Python for-loops), where it is furthermore possible to make use of OpenMP\footnote{\url{http://openmp.org/}} for automatic parallelisation of the code to improve execution times on machines with many CPU cores. The second improvement is algorithmic in nature and aims to speed-up the geometric part of the computation: the geodesics calculation and height profile generation, which is preparatory work before the actual P.452-algorithm is employed. A path from the map centre to a pixel on the edge of the map crosses many other pixels in-between. We calculate all geodesics between map edge pixels and the centre and use a caching technique to identify the edge-geodesic which comes closest to the pixel centre for each interior pixel. This essentially transforms the problem from $O(n^2)$ to $O(n)$. Unfortunately, the P.452 algorithm itself can not be treated with a similar technique, because the path elevation angles (and thus also if a path is line-of-sight or trans-horizon) are unique for each pixel as they depend on the terrain height profile between the pixel and the map centre.

\begin{figure}[t]
\includegraphics[width=8.3cm]{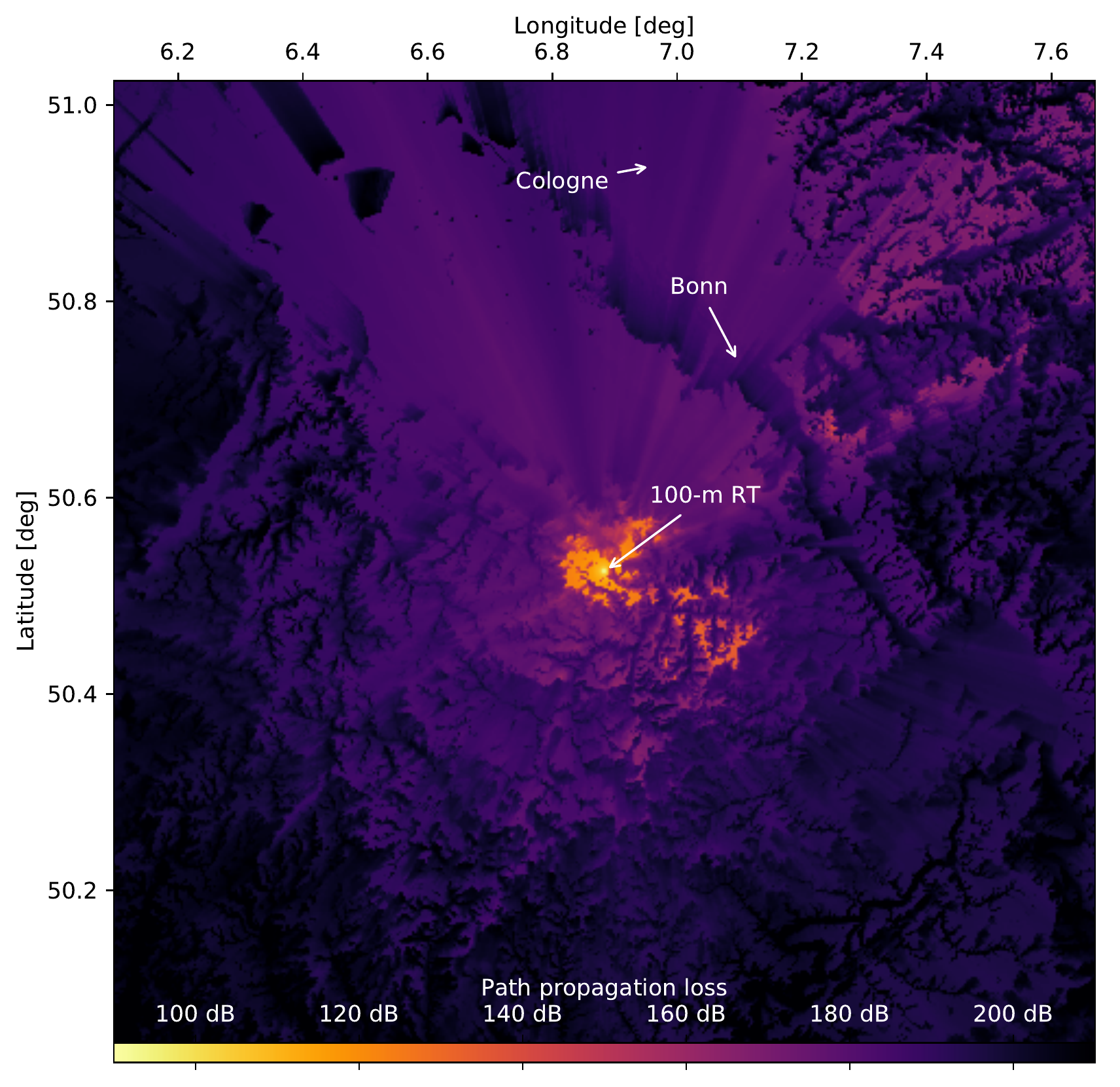}
\caption{Example map of path propagation losses around the Effelsberg observatory. Based on NASA SRTM data \citep{farr07}.}
\label{fig:attenmap}
\end{figure}

Using the attenuation map-making functionality is again a two-step process. First, a helper object is created that caches the terrain and path parameters, which is then fed into the P.452 algorithm:
\begin{lstlisting}[frame=,caption=Fast path propagation loss map-making.]
import astropy.units as u
import pycraf.conversions as cnv
import pycraf.pathprof as pp

pp.SrtmConf.set(download='missing')

# Effelsberg radio telescope site
lon_rt = 6.88361 * u.deg
lat_rt = 50.52483 * u.deg

# Map parameters
msize_lon, msize_lat = (1, 1) * u.deg
map_resolution = 3 * u.arcsec
zone_t = pp.CLUTTER.URBAN
zone_r = pp.CLUTTER.SUBURBAN

# P.452 parameters
frequency = 10 * u.GHz
temperature = 290 * u.K
pressure = 1013 * u.hPa
time_percent = 2 * u.percent
h_rt, h_tx = (50, 20) * u.m


hprof_cache = pp.height_map_data(
    lon_rt, lat_rt,
    msize_lon, msize_lat,
    map_resolution=map_resolution,
    zone_t=zone_t, zone_r=zone_r,
    )

results = pp.atten_map_fast(
    frequency,
    temperature,
    pressure,
    h_rt, h_tx,
    time_percent,
    hprof_cache,
    )

total_atten = results['L_b']
\end{lstlisting}
Again, we omit the lengthy plot code, but the result of this is visualized in Fig.~\ref{fig:attenmap}. The \texttt{hprof\_cache} object is a Python \texttt{dict}-like object containing the cached parameters and could also be edited by the user before calling \texttt{atten\_map\_fast}. A typical example of such editing would be the assignment of a different clutter zone to each of the pixels -- given that the user has access to a suitable database containing the true clutter types.

The \texttt{pathprof} sub-package also provides a large amount of other utility functions, e.g., to plot terrain maps or export maps to a Geographic Information System (GIS) file format such as the Keyhole Markup Language (KML); see online documentation for further information.

\subsection{Protection thresholds}

Since \texttt{pycraf} was developed for compatibility studies involving radio astronomy, the current \texttt{protection} sub-package only offers thresholds relevant in that context. Most importantly, this covers the tables with protection thresholds given in \citetalias{itu_ra769_2}:
\begin{lstlisting}[frame=,caption=\citetalias{itu_ra769_2} thresholds.]
from pycraf import protection

tab = protection.ra769_limits(
    mode='continuum'
    )

print(tab[[
    'frequency', 'bandwidth',
    'T_A', 'T_rx',
    'Plim'
    ]])

# frequency bandwidth  T_A  T_rx  Plim
#    MHz       MHz      K    K   dB(W)
# --------- --------- ----- ---- ------
#        13         0 50000   60 -184.6
#        26         0 15000   60 -187.9
#        74         2   750   60 -195.0
#       152         3   150   60 -199.5
#       325         7    40   60 -201.0
#       408         4    25   60 -202.9
#       ...       ...   ...  ...    ...
#     23800       400    15   30 -195.6
#     31550       500    18   65 -192.4
#     43000      1000    25   65 -190.6
#     89000      8000    12   30 -189.4
#    150000      8000    14   30 -189.2
#    224000      8000    20   43 -187.6
#    270000      8000    25   50 -186.8
# Length = 21 rows
\end{lstlisting}
Values from this table can then subsequently be used to compare received power levels of a potential interferer with the permitted power levels. Almost all compatibility studies that involve RAS refer to the RA.769 methodology or thresholds in one form or another.

\subsection{Antenna patterns}\label{subsec:pycraf_antenna}
In many realistic compatibility studies, the antenna patterns of transmitter and receiver have to be accounted for. This can be more complex than what might be naively assumed. For example, the base stations and partly also the user equipment of the next-generation mobile communications standard 5G will utilize phased array antennas. To calculate the effective gain for the propagation path, one must account for the direction of the formed beam, the orientation (and rotation) of the antenna frame normal vector, and the relative position of the propagation path with respect to the antenna frame. We will return to this in Section~\ref{sec:5g}.

\begin{figure}[t]
\includegraphics[width=8.3cm]{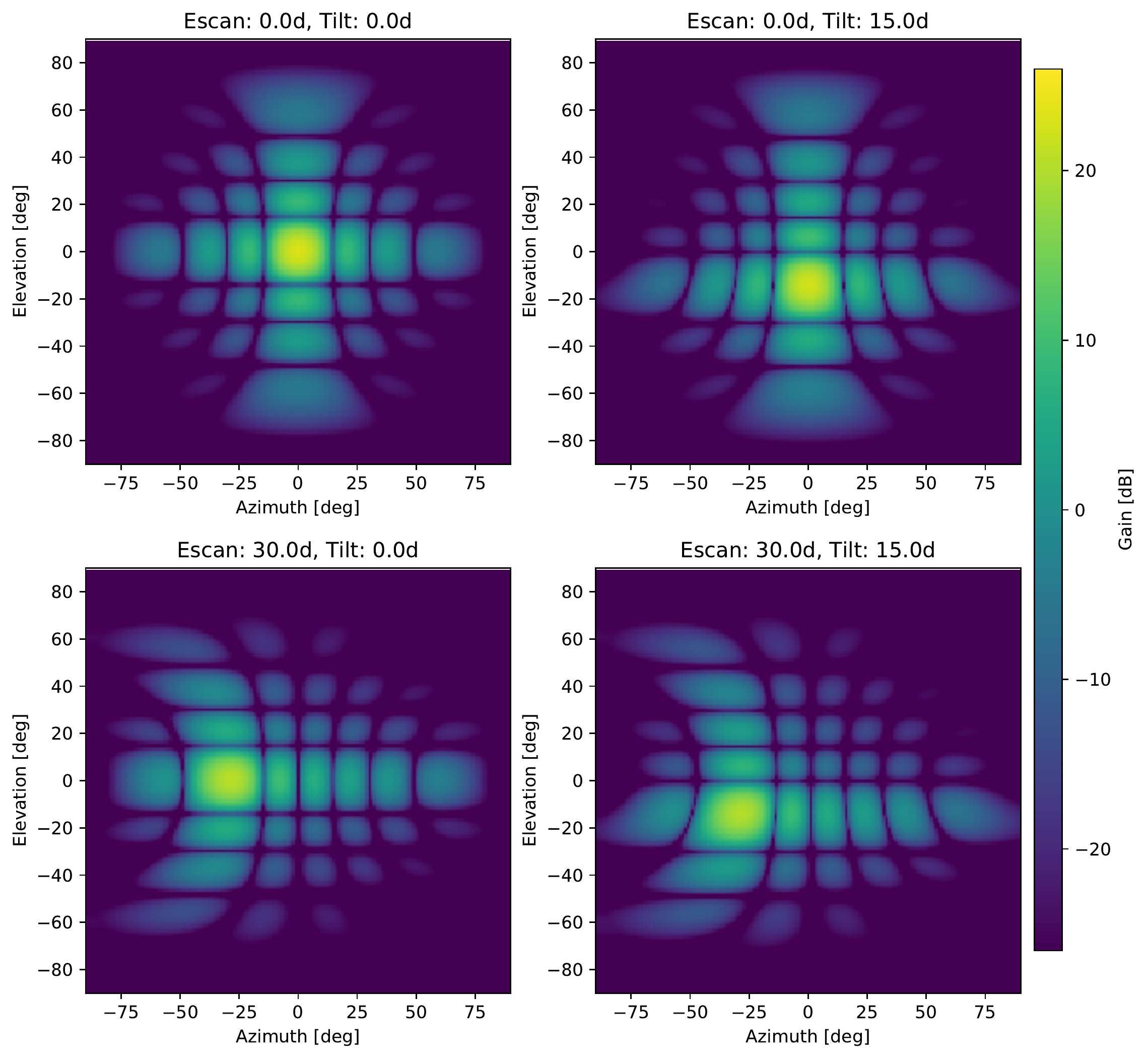}
\caption{Composite phased-array antenna patterns of the next mobile-communications standard 5G as defined in \citetalias{itu_m2101_0}.}
\label{fig:compositepatterns}
\end{figure}

In \texttt{pycraf} several (simplified) antenna models are provided that implement various ITU-R recommendations, which shall be used for spectrum management studies. As an example, the fairly complex 5G composite antenna patterns can be obtained in the following manner:
\begin{lstlisting}[frame=,caption=5G composite antenna patterns.]
import numpy as np
import astropy.units as u
import pycraf.conversions as cnv
import pycraf.antenna as ant

azims = np.arange(-180, 180) * u.deg
elevs = np.arange(-90, 90) * u.deg

# antenna parameters according to
# ITU-R Rec. M.2101
G_Emax = 5 * cnv.dB
A_m, SLA_nu = (30, 30) * cnv.dimless
azim_3db, elev_3db = (65, 65) * u.deg
d_H, d_V = (0.5, 0.5) * cnv.dimless
N_H, N_V = 8, 8

azim_i, elev_j = (0, 0) * u.deg

gains = ant.imt2020_composite_pattern(
    azims[np.newaxis],
    elevs[:, np.newaxis],
    azim_i, elev_j,
    G_Emax,
    A_m, SLA_nu,
    azim_3db, elev_3db,
    d_H, d_V,
    N_H, N_V,
    )
\end{lstlisting}
where the \texttt{azim\_i} and \texttt{elev\_j} angles define the pointing of the formed beam in the antenna frame coordinates (azimuth and elevation). In Fig.~\ref{fig:compositepatterns} examples are shown for four different beam directions: $(0^\circ, 0^\circ)$, $(30^\circ, 0^\circ)$, $(0^\circ, 15^\circ)$, and $(30^\circ, 15^\circ)$.







\subsection{Atmospheric attenuation}
Once the propagation path exceeds a certain length, e.g., for communication links to satellites or for radio-astronomical observation of extra-terrestrial sources, the attenuation by the gases in the atmosphere of the Earth becomes important, especially at higher frequencies. Again, there exists a ITU-R recommendation, P.676,
which was implemented in the \texttt{atm} sub-package. It can be used to predict the attenuation as a function of frequency, path geometry, and atmospheric parameters. The two molecules that play a major role are oxygen (so-called dry component) and water (wet component), both introducing spectral-line emission and absorption. Furthermore,  oxygen also produces the continuous non-resonant Debye-spectrum (below 10~GHz). At frequencies above 100~GHz continuous attenuation by nitrogen starts to play a role as well.

The calculation of atmospheric attenuation is performed via a ray-tracing approach. Let us assume a radio signal enters Earth's atmosphere and travels to a receiving station close to the ground. First, the atmosphere is divided into layers (several hundreds or more) and for each of these the attenuation of the incident signal is calculated, such that the signal, which leaves a layer is somewhat weaker than the signal that entered a layer. At the same time, the layer itself emits radio waves having the aforementioned spectral properties. Each layer has different physical properties, such as temperature, pressure, water content, and refractivity. These ``height profiles'' have to be known to calculate the specific attenuation per layer. For ITU-R related studies, it is widely accepted to work with mean profiles, which were averaged over a year. The ``standard profile'' is provided in \citetalias{itu_p835_5}
along with five more specialised profiles, associated with the geographic latitude and season (``high-latitude summer/winter'', ``mid-latitude summer/winter'', and ``low latitude''). For example:
\begin{lstlisting}[frame=,caption=Atmospheric attenuation.]
import numpy as np
import astropy.units as u
import pycraf.conversions as cnv
from pycraf import atm

# define height grid
height_grid = np.arange(0, 85, 0.1) * u.km

# query profile_standard function
(
    temperatures,
    pressures,
    rho_water,
    pressures_water,
    ref_indices,
    humidities_water,
    humidities_ice
    ) = atm.profile_standard(height_grid)
\end{lstlisting}



\begin{figure}[t]
\includegraphics[width=8.3cm]{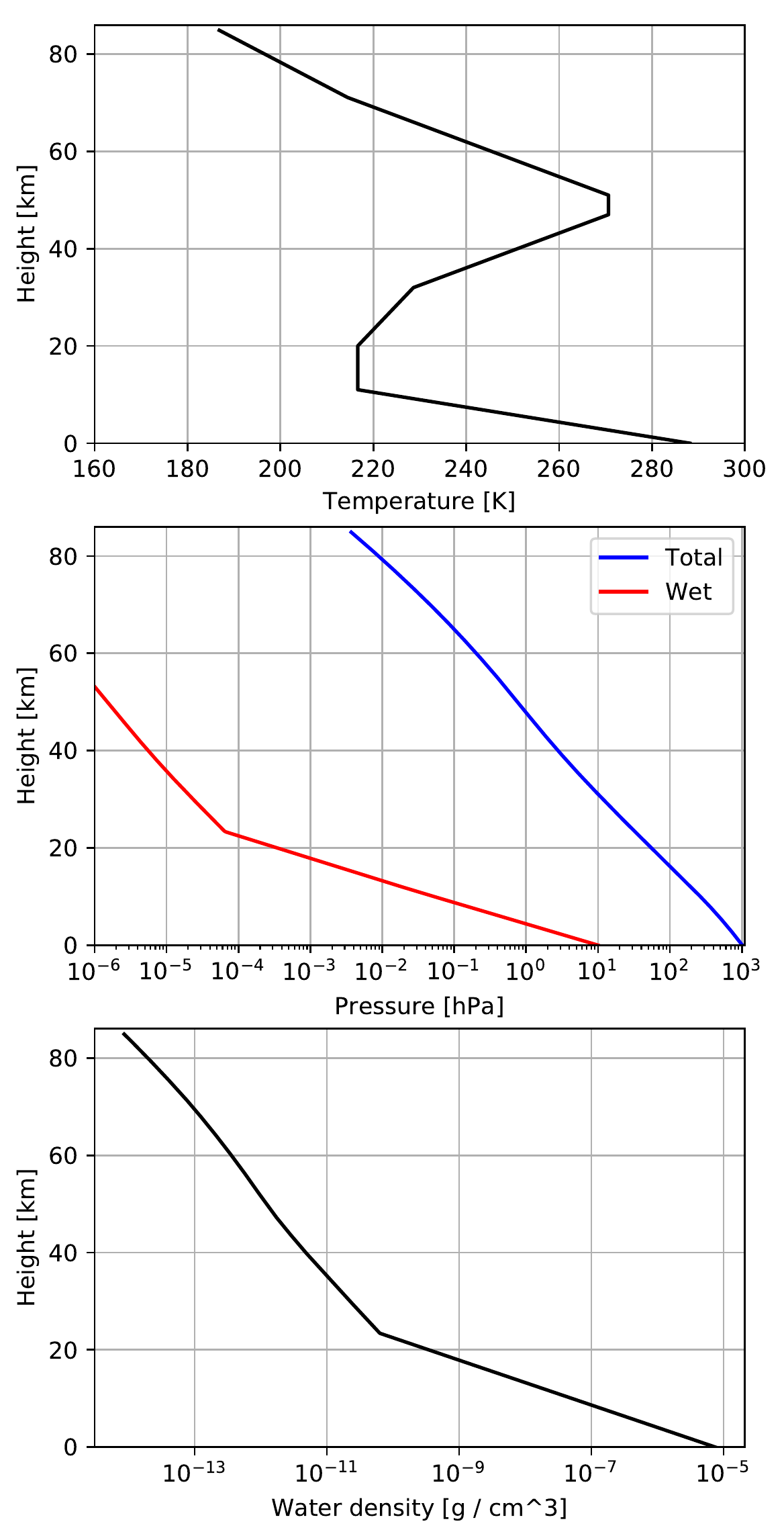}
\caption{Standard atmospheric height profiles (temperature, pressure, and water density) according to \citetalias{itu_p835_5}.}
\label{fig:atm_heightprof}
\end{figure}
The resulting profiles are displayed in Fig.~\ref{fig:atm_heightprof}. \citetalias{itu_p676_10} contains tables with the resonance lines of oxygen and water, necessary to calculate the specific attenuation for one layer (or propagation paths close to the ground, which stay in a single layer). The function \texttt{atten\_specific\_annex1} can be used for that:
\begin{lstlisting}[firstnumber=last,frame=]
freqs = np.arange(1, 1000) * u.GHz
p_total = 1013 * u.hPa
temperature = 290 * u.K
rho_water = 7.5 * u.g / u.m ** 3

# need dry and wet pressure separately:
p_wet = atm.pressure_water_from_rho_water(
    temperature, rho_water
    )
p_dry = p_total - p_wet

print('O2 pressure: {:.2f}'.format(p_dry))
print('H2O vapor partial pressure: '
      '{:.2f}'.format(p_wet))
# Oxygen pressure: 1002.96 hPa
# Water vapor partial pressure: 10.04 hPa

a_dry, a_wet = atm.atten_specific_annex1(
    freqs, p_dry, p_wet, temperature
    )
\end{lstlisting}

\begin{figure}[t]
\includegraphics[width=8.3cm]{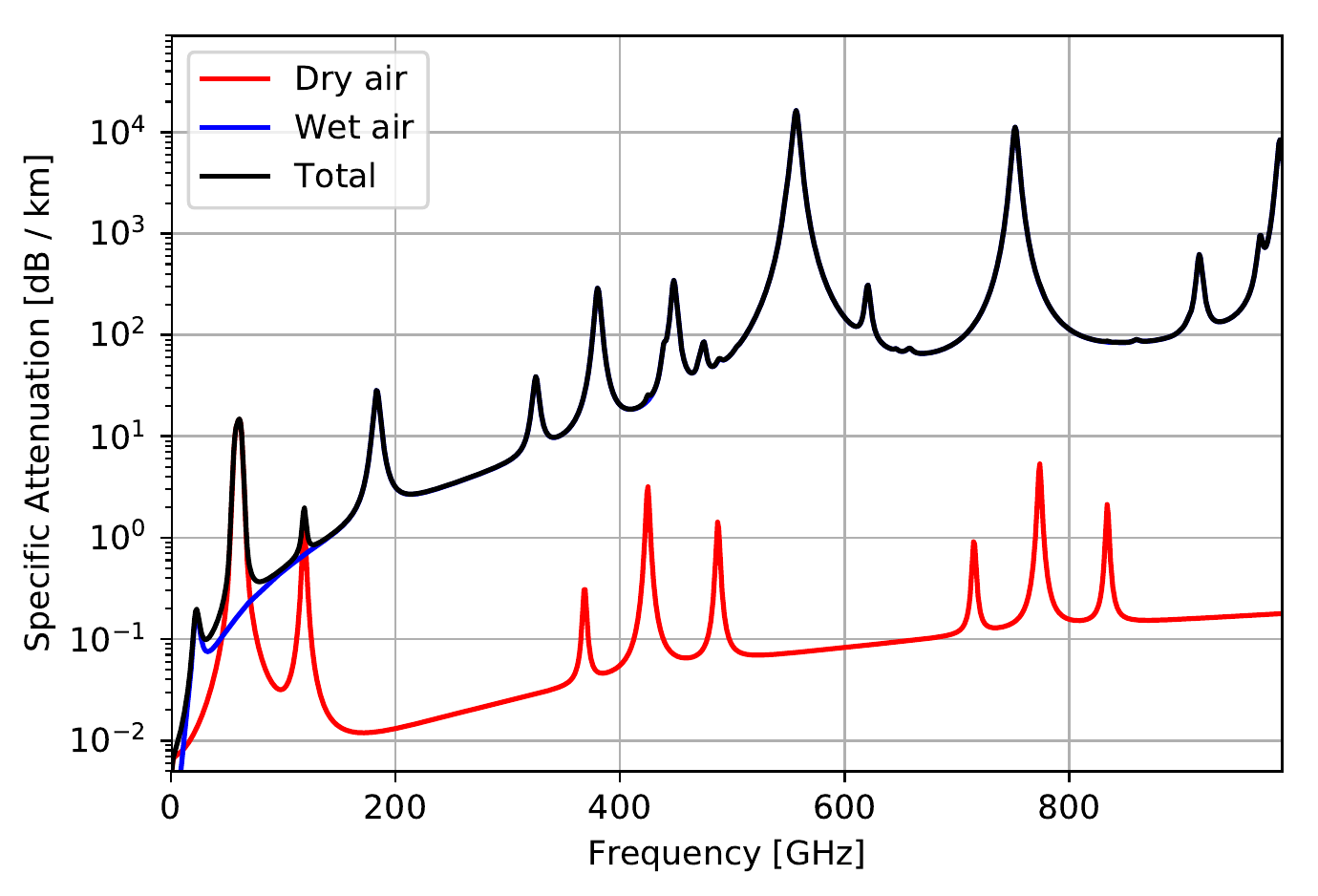}
\caption{Specific attenuation for a layer with a total pressure of 1013~hPa, a temperature of 290~K, and a specific water density of $7.5~\mathrm{g\,cm}^{-3}$ according to \citetalias{itu_p676_10}.}
\label{fig:atm_specific_atten}
\end{figure}
The resulting spectrum is shown in Fig.~\ref{fig:atm_specific_atten}. Calculating such spectra for each layer allows us to perform the ray-tracing for a slant path through the full atmosphere (with the function \texttt{atten\_slant\_annex1}):
\begin{lstlisting}[firstnumber=last,frame=]
elevation = 15 * u.deg
station_altitude = 300 * u.m
freqs = np.arange(1, 100, 0.1) * u.GHz

(
    total_atten, refraction, tebb
    ) = atm.atten_slant_annex1(
        freqs, elevation,
        station_altitude,
        atm.profile_highlat_winter
        )
\end{lstlisting}







\begin{figure}[t]
\includegraphics[width=8.3cm]{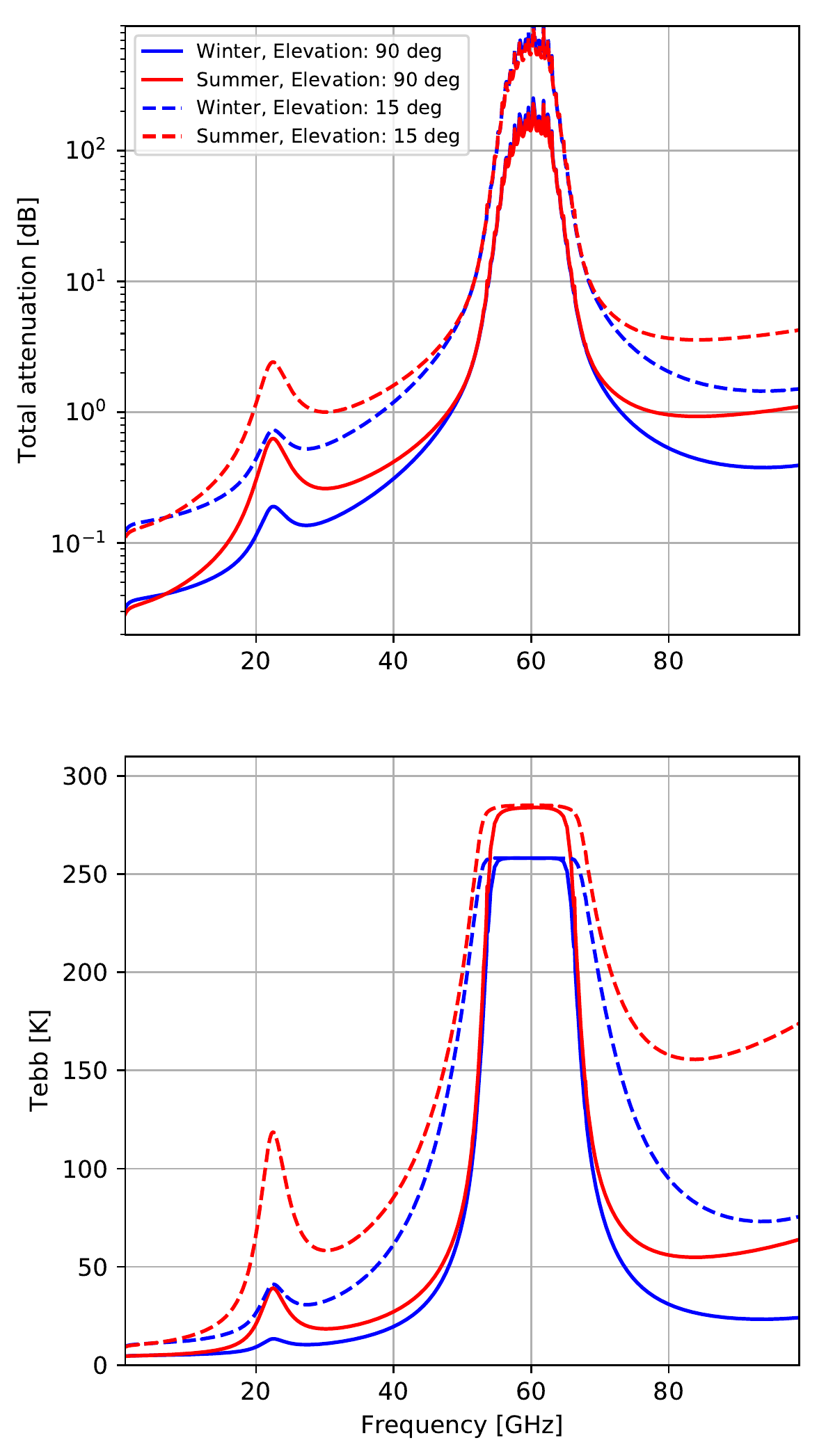}
\caption{Atmospheric attenuation of a slant path for Summer/Winter high-latitude standard profiles according to \citetalias{itu_p676_10}. The lower panel shows the equivalent-black-body temperature of the full atmosphere.}
\label{fig:atm_slant_atten}
\end{figure}
In Fig.~\ref{fig:atm_slant_atten} the total atmospheric attenuation of a slant path for Summer/Winter high-latitude standard profiles is presented, along the so-called equivalent-black-body temperature, $T_\mathrm{ebb}$, of the full atmosphere with major contributions by the atmospheric emission itself.

\subsection{Satellite two-line elements}

For studies involving a satellite or a satellite network it is often necessary to determine their position relative to an Earth station. Satellite orbits vary slowly in time as a consequence of the gravitational influence of planetary system bodies, most importantly the Moon, the Sun, and Jupiter. Furthermore, the orbits are subject to drag forces induced by the outer layers of the atmosphere, especially, in the case of low-earth orbit (LEO) satellites. Satellites are therefore  constantly monitored by stations on Earth to update their orbital parameters. These are provided in a standardized format, the so-called two-line element (TLE) sets, which are usually updated once per day and can be retrieved from the Space Track\footnote{\url{https://www.space-track.org}} or CelesTrak\footnote{\url{http://celestrak.com/}} websites. As an example, the TLE of the International Space Station (ISS) for September, 20, 2008 looks like this:
\begin{lstlisting}[numbers=none,frame=,caption=,basicstyle=\tiny]
ISS (ZARYA)
1 25544U 98067A   08264.51782528 -.00002182  00000-0 -11606-4 0  2927
2 25544  51.6416 247.4627 0006703 130.5360 325.0288 15.72125391563537
\end{lstlisting}
The meaning of the entries is described e.g. on Wikipedia\footnote{\url{https://en.wikipedia.org/wiki/Two-line_element_set}}. The fields 7 and 8 (line 1, columns 19--20 and 21-32), i.e., \texttt{08} and \texttt{264.51782528} encode the epoch (year and day of year) at which the other orbital parameters are valid. From this point in time, one can then \textit{propagate} the orbit.

To account for the aforementioned distortions of orbits, the calculations behind the propagation algorithm are rather complex and \texttt{pycraf} internally uses the Python package \texttt{sgp4} for this. Provided with a TLE string, \texttt{sgp4} can calculate the position of the satellite in the Geocentric reference frame for any point in time. However, if the time difference to the TLE epoch is significantly larger than a day, the result may be completely wrong.

For compatibility studies one usually needs the satellite position in the antenna reference frame of the ground station. Therefore, the \texttt{satellite} sub-package of \texttt{pycraf}, which is otherwise just a thin wrapper around \texttt{sgp4}, adds some routines for this coordinate conversion. One only needs to specify the coordinates of the observatory and the desired time:
\begin{lstlisting}[frame=,caption=Satellite TLEs.]
from astropy import coordinates, time
from pycraf import satellite as sat

tle_string = '''ISS (ZARYA)
1 25544U 98067A   08264.51782528 ...
2 25544  51.6416 247.4627 0006703...'''

# define observer location
loc = coordinates.EarthLocation(
    6.88375, 50.525, 366.
    )

# create a SatelliteObserver instance
sat_obs = sat.SatelliteObserver(loc)

obstime = time.Time(
    '2008-09-20T20:00:00.000000',
    format='isot', scale='utc'
    )

az, el, dist = sat_obs.azel_from_sat(
    tle_string, obstime
    )
print('Azimuth  : {:6.1f}'.format(az))
print('Elevation: {:6.1f}'.format(el))
print('Distance : {:6.1f}'.format(dist))
# Azimuth  :   80.1 deg
# Elevation:   11.1 deg
# Distance : 1257.7 km
\end{lstlisting}







\begin{figure}[t]
\includegraphics[width=8.3cm]{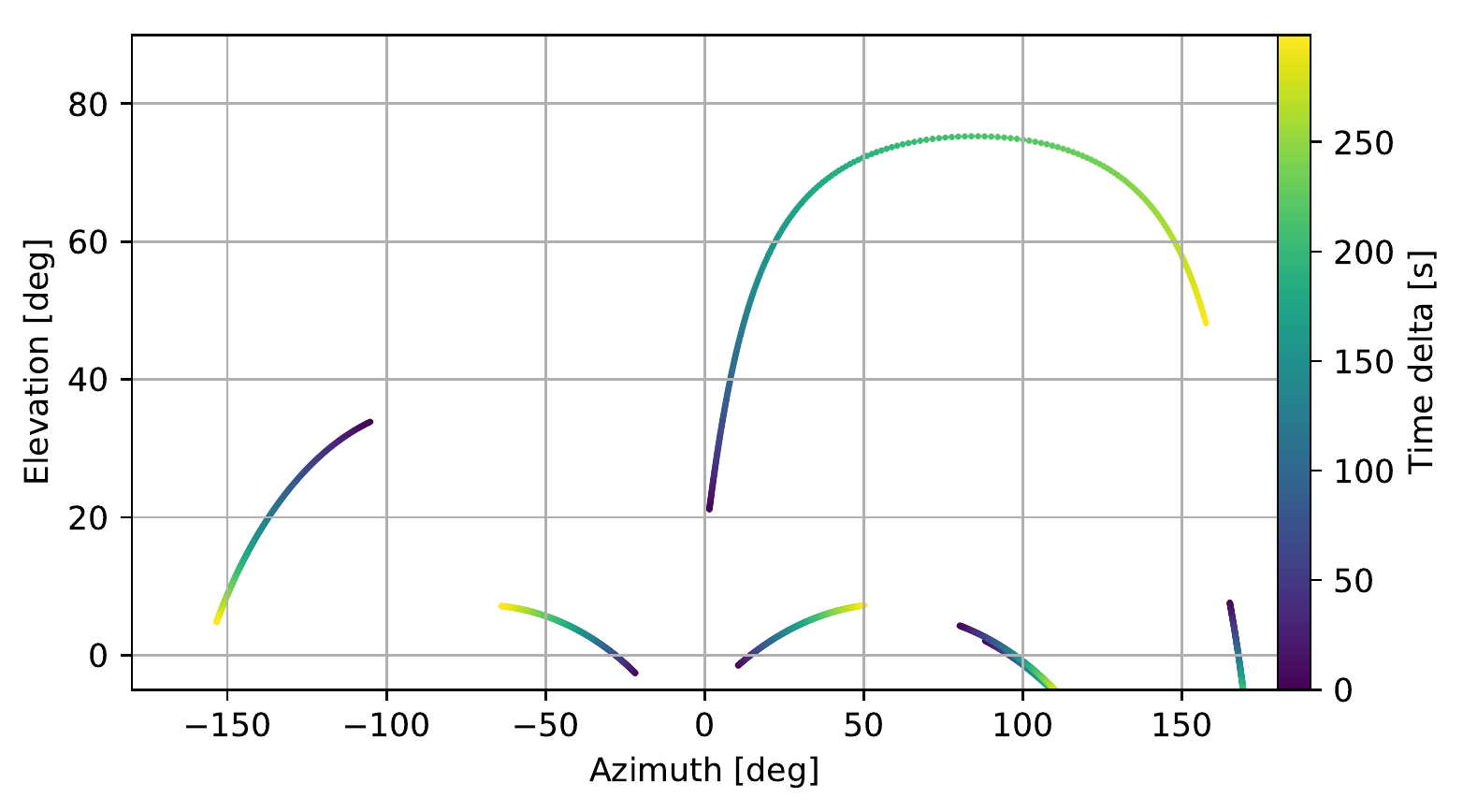}
\caption{Position of visible IRIDIUM network satellites in the topocentric reference frame for an observer at the Effelsberg observatory. Displayed are the azimuth and elevation of the satellites for five minutes starting from January, 4, 2018, 21:45 (UTC).}
\label{fig:iridium_sats}
\end{figure}
As another example, Fig.~\ref{fig:iridium_sats} shows the position of the IRIDIUM network satellites, as seen from an observer at the Effelsberg observatory. IRIDIUM satellites are LEOs and move fast over the visible sky. To visualize this, the azimuth and elevation of the satellites were computed for a time range of five minutes.

\section{Compatibility of a 5G cell phone network and RAS}\label{sec:5g}
With the \texttt{pycraf} package introduced in Section~\ref{sec:pycraf} it is very easy to do simple compatibility studies. Especially, for a point-to-point analysis, there is not much more to it than the determination of the emitted power and the path loss and finally the comparison of  the received power with the compatibility threshold level. These steps could easily be put into a stand-alone software package, perhaps with a graphical user interface. A need for additional functionality may however arise sometimes, or one may be faced with more complex scenarios right from the beginning. Therefore, \texttt{pycraf} was designed as a programming library. Although this makes the simplest projects slightly more complicated (because one always need to write some Python code), it is well-suited to deal with more challenging studies. One of these more complicated cases is discussed in this section.

\subsection{Context}
The mobile telecommunication industry wants to utilize more bandwidth for their next-generation mobile communications standard, which is called 5G.
At the last world  radio communications conference (WRC), an agenda item (AI 1.13) was proposed for the next WRC in 2019. It is entitled ``Work plan and proposed liaison to Task Group 5/1 on spectrum needs for the terrestrial component of IMT\footnote{International Mobile Telecommunications} in the frequency range between 24.25 GHz and 86 GHz''. Although there is currently some tendency among vendors and administrations to favour the lower part of the frequency range mentioned in the agenda item (the 24.25--27.5~GHz sub-band)  we are required to study all possible sub-bands without a predetermined conclusion for AI 1.13.

The Committee on Radio Astronomy Frequencies (CRAF) -- an expert committee of European radio astronomers involved in spectrum management -- is submitting input documents with compatibility studies to various regional and global working groups (ITU-R TG\,5/1, ECC~PT\,1). These studies  attempt to determine the compatibility of the next-generation 5G equipment with radio-astronomical observations. Depending on the designated 5G band, protected RAS bands would in some cases be located in the out-of-band (OOB) or spurious domain with respect to the IMT carrier frequency, in other cases we may face the in-band sharing scenario, where both, Tx and Rx, frequency ranges overlap. The latter case usually demands larger separation distances between interfering stations (the IMT equipment) and the victim station (the radio telescope) to ensure that radio-astronomical observations do not suffer interference. Because of
 the additional attenuation of the emitted power a compatibility is more  easily obtained in the OOB or spurious cases.

Here, we describe how the compatibility calculations for one of these studies were made, focussing on the 24.25--27.5~GHz frequency band. RAS has a primary allocation for 23.6--24~GHz, hence, this is an example for a OOB-domain scenario. For compatibility studies on the ITU level, one often analyses the ``generic'' case, where no terrain height (above the Earth ellipsoid) is assumed, so that the results do not depend on the terrain details of specific sites. Of course, RAS stations in mountainous regions may have some level of natural shield from surrounding hill tops, so that generic-study results are only indicative. Still, many RAS stations are located in regions with relatively flat terrain, e.g., the Westerbork Synthesis Radio telescope in the Netherlands.

\textit{We emphasize that the results presented here sensitively depend on the chosen transmitter properties and deployment densities. The assumptions and values, which are used in the following, are preliminary, as some of relevant (ITU-R) recommendations and guidelines are not yet finalized. Therefore, one should be careful with drawing conclusions from our study.}

\subsection{Setup I: Transmitter and receiver properties}\label{subsec:setup1_5g}

5G networks will likely be ``classic'' mobile networks, consisting of base stations (BS) and user equipment (UE), although alternatives such as mesh-network based topologies seem also viable. Both, BS and UE feature phased-array antenna patterns, as described in Section~\ref{subsec:pycraf_antenna}. The transmitter properties (to be used for compatibility calculations) are summarized in \citetalias{itu_tg5_1_doc36_att2}. For BS, $8\times8$ individual antenna elements, each having a power output of 10~dBm per carrier (before Ohmic losses), are foreseen. UE will use smaller arrays, of $4\times4$ elements, but with the same power levels. For the spurious domain, a spectrum block edge mask of $-$38~dBc per carrier bandwidth (200~MHz) for BS and $-$32~dBc/carrier for UE is in place, such that the spectral power density within the RAS band is $-$13~dBm/MHz in both cases. For UE an additional so-called body loss of 4~dB has to be accounted for, because in many situations a person holding the user device (e.g., a smart phone) will provide  additional shielding. Integrating the spectral power over the RAS band yields a total of 13~dBm.

\begin{figure}[t]
\includegraphics[width=8.3cm]{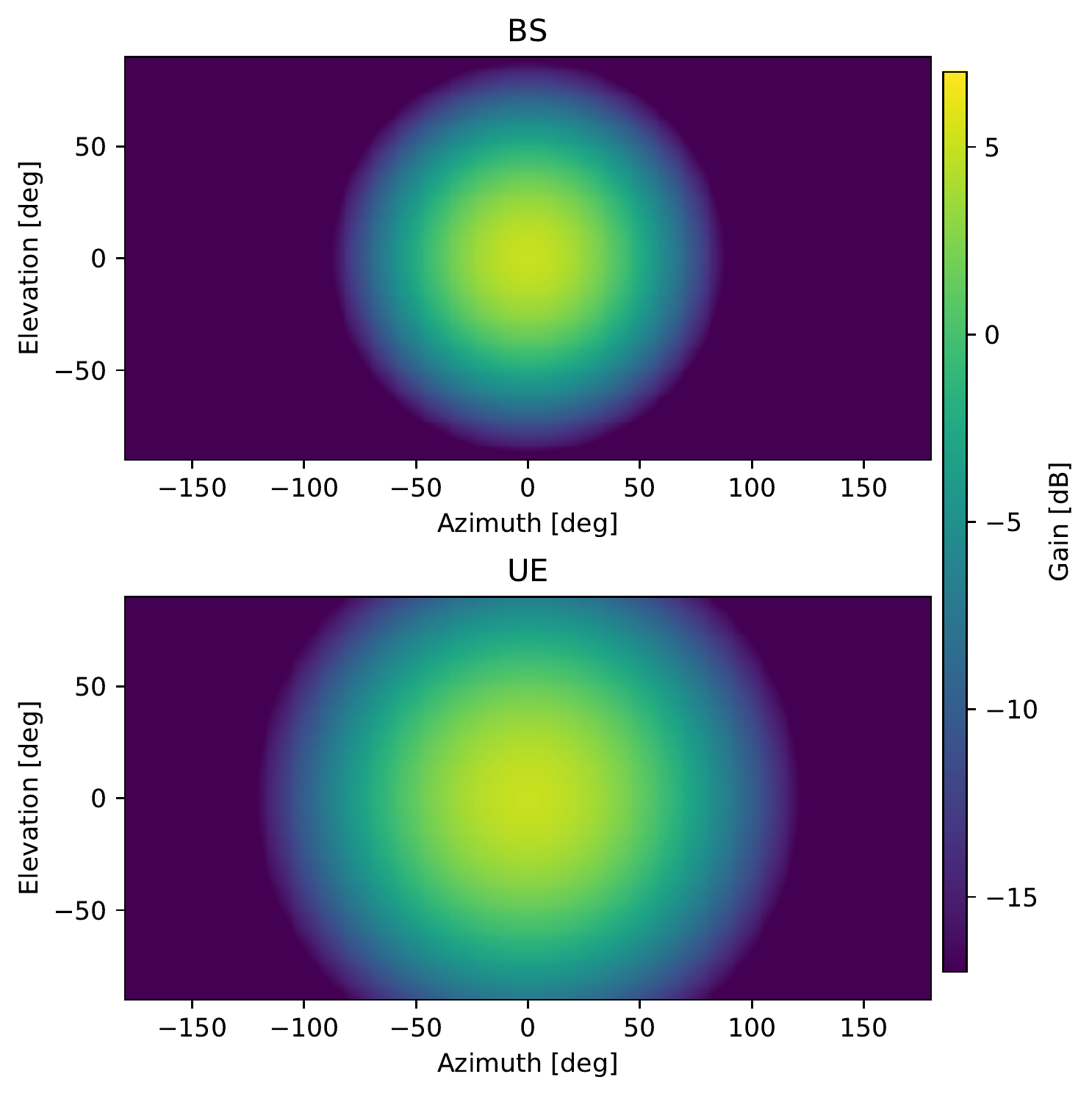}
\caption{Single-element antenna patterns of the next mobile-communications standard 5G as defined in \citetalias{itu_m2101_0}.}
\label{fig:singlepattern}
\end{figure}

\citetalias{itu_tg5_1_doc36_att2} makes a distinction for different zone types, \textit{urban}, \textit{suburban} and \textit{suburban open-space} regions with different number densities and antenna heights. BS antenna heights will be 15~m for suburban open-space zones and 6~m for the rest, while UE is always at 1.5~m height. Even though the 5G phased array antennas allow dynamic beam forming, \citetalias{itu_m2101_0} advises to use the single-element patterns for studies in the spurious domain. We follow this advice here, although there are diverging opinions about it as some consider it unlikely that the signal phasing should be completely inefficient close to the carrier frequency.  Fig.~\ref{fig:singlepattern} shows the resulting antenna gain patterns for BS and UE, respectively. The composite antenna patterns will still be required (see Section~\ref{subsec:aggregation}), because they influence the link budget (or coupling loss) between BS and UE, which in turn determines the level of power control in the user devices -- UE can increase or decrease the transmitted power based on the coupling loss for more efficient battery use.

On the receiver side, an isotropic antenna (0~dBi) is usually assumed for these kinds of studies, with a height of 50~m above ground, which would be typical for a 100-m class radio telescope. \citetalias{itu_ra769_2} permits an interference power level of $-$165~dBm (in continuum mode). Therefore, the minimal coupling loss (MCL), i.e., the attenuation induced by the path propagation loss plus the gain or attenuation due to the Tx antenna gain, must be at least 178~dB. The MCL ultimately determines the minimal distance (also called separation distance) that a device must have from the Rx to avoid problems with RAS observations.

\subsection{Path propagation loss}\label{subsec:pathloss_5g}
With the functionality contained in \texttt{pycraf}, it is very easy to determine the path loss as a function of distance between Tx and Rx. Here, a distinction is to be made between the sub-/urban and suburban open-space areas, because the antenna height plays a  role for the propagation. The default clutter model of \citetalias{itu_p452_16} is not be applied in studies related to agenda item AI~1.13, instead the use of the model provided in \citetalias{itu_p2108_0} is mandatory. It depends only on frequency, distance, and a quantity called location percentage, $p_\mathrm{L}$, which is the percentage of all emitters producing the lowest clutter loss. For example, if $p_\mathrm{L}=2\%$, the returned value, $L_\mathrm{clutter}$, indicates that for 2\% of all devices the actual clutter loss will fall short of $L_\mathrm{clutter}$. In other words, P.2108 provides the cumulative distribution function for the clutter. We also note that the distance dependence in \citetalias{itu_p2108_0} is only relevant for very small values below $\sim$2~km.

\begin{figure}[t]
\includegraphics[width=8.3cm]{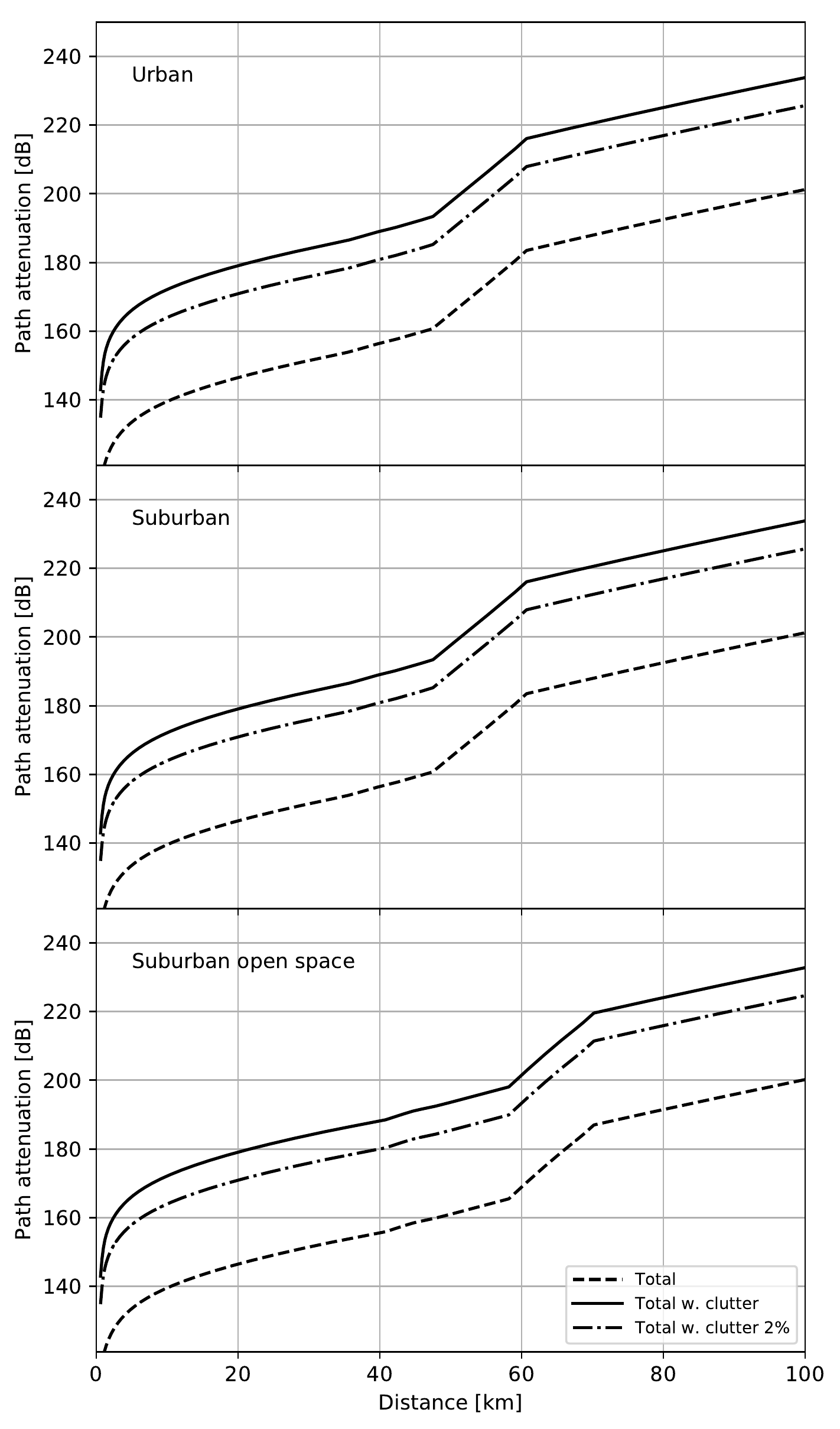}
\caption{Path propagation loss for BS at 24~GHz according to \citetalias{itu_p452_16} (dashed line). The black solid line includes the median clutter loss predicted by \citetalias{itu_p2108_0}, the dash-dotted line was computed for a location percentage of $p_\mathrm{L}=2\%$ (see text).}
\label{fig:pathloss_bs}
\end{figure}

\begin{figure}[t]
\includegraphics[width=8.3cm]{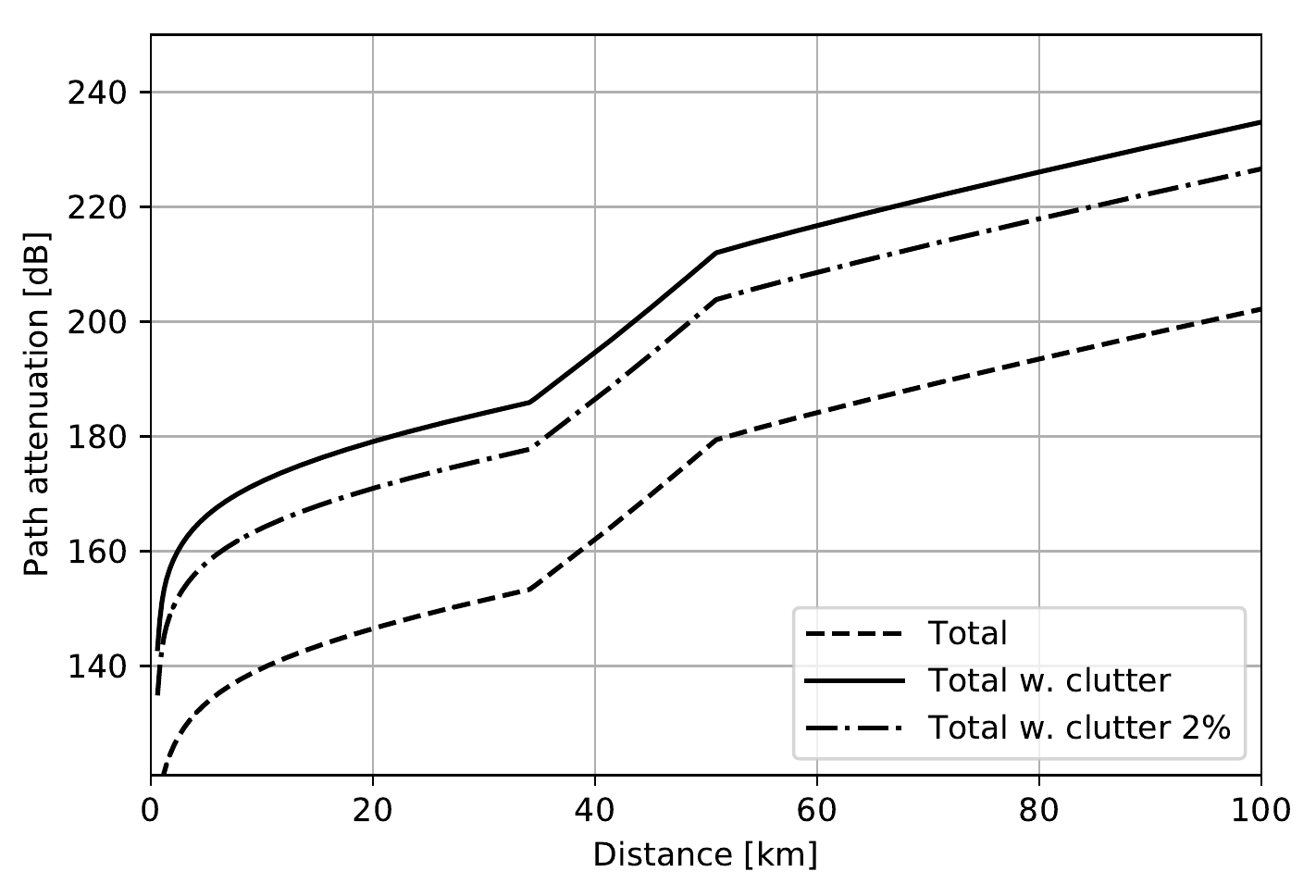}
\caption{As Fig.~\ref{fig:pathloss_bs} for UE.}
\label{fig:pathloss_ue}
\end{figure}

In Figs.~\ref{fig:pathloss_bs} and \ref{fig:pathloss_ue} the resulting path propagation loss for BS and UE is displayed. The rapid increase of the path loss between about 35 and 50~km is caused by the geometry of the path. There, the line-of-sight path (small distances) turns into a trans-horizon path, meaning that diffraction kicks in.

\subsection{Single-interferer scenario}
The single-interferer scenario is an extreme case, where only one IMT device is considered having maximally possible antenna gain (i.e., the antenna shall be directed towards the RAS station\footnote{In case of BS, the array antennas are installed with a down-tilt of $-10^\circ$, so that the single-element pattern maximum is usually not pointing towards the RAS station.}) and very low $p_\mathrm{L}$ (which means the clutter loss will be unusually small). In reality, such a situation will rarely occur, but it is still valuable to study this case, because it already gives a rough indication about the magnitude of the necessary separation distances. Very often, the single-interferer separation distances define a zone, where operation of the interfering service will be critical.

\begin{figure}[t]
\includegraphics[width=8.3cm]{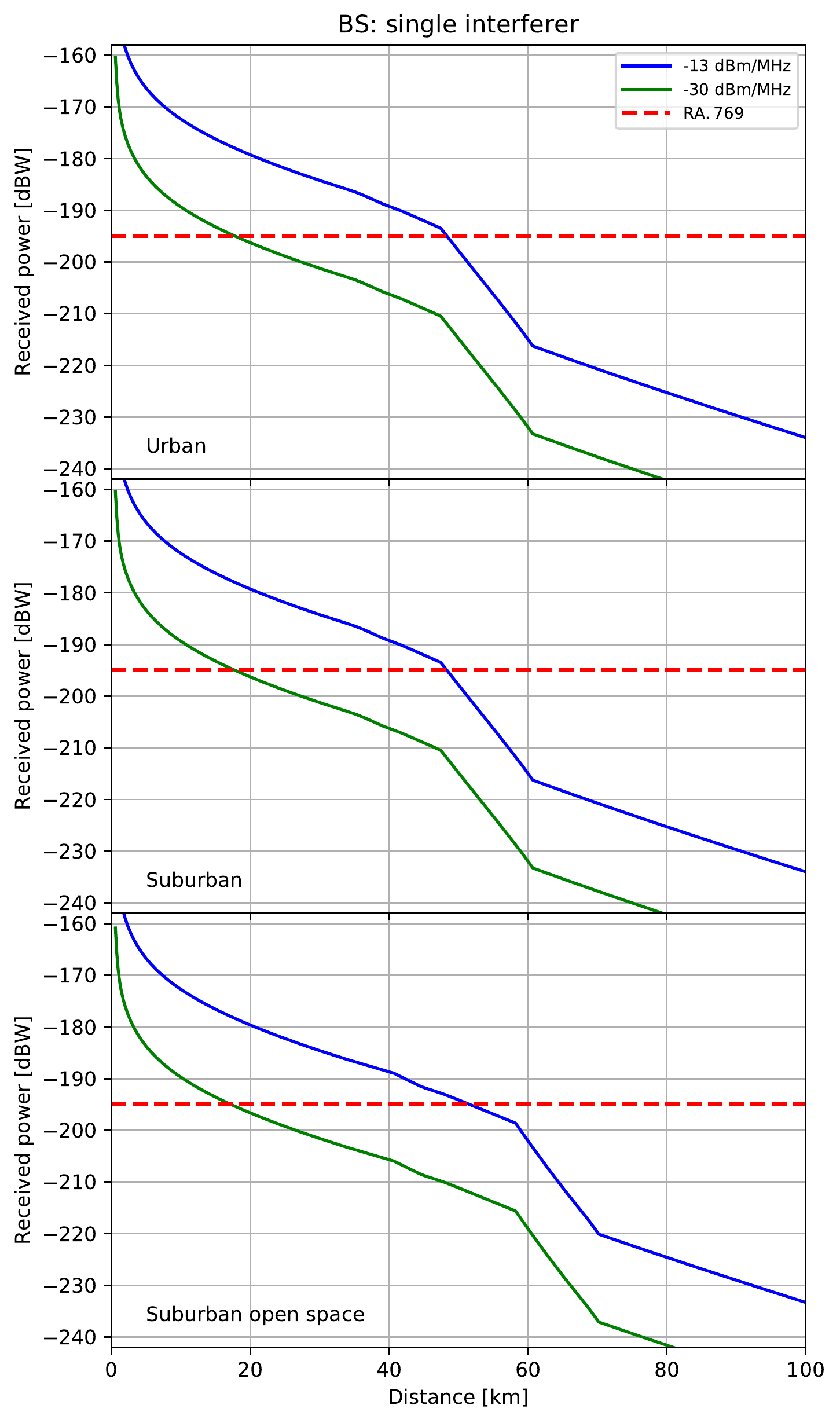}
\caption{Received power levels of a BS for the worst-case single interferer scenario. Two different spurious-domain Tx power levels were assumed. The red dashed line is the RA.769 threshold, which should not be exceeded.}
\label{fig:singleinterferer_bs}
\end{figure}

\begin{figure}[t]
\includegraphics[width=8.3cm]{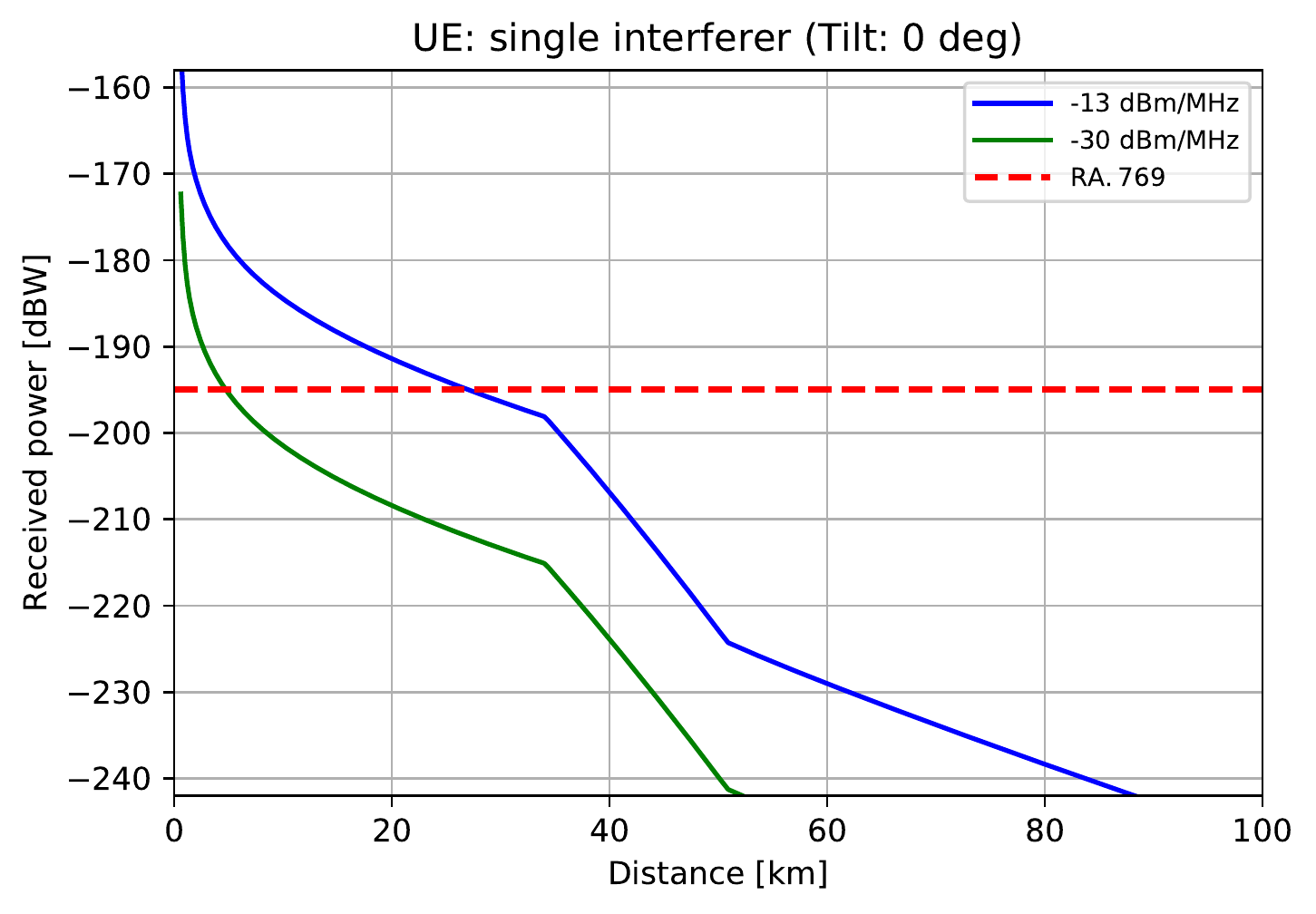}
\caption{As Fig.~\ref{fig:singleinterferer_bs} for UE.}
\label{fig:singleinterferer_ue}
\end{figure}

Figures~\ref{fig:singleinterferer_bs} and \ref{fig:singleinterferer_ue} contain the resulting power levels received from a single BS or UE with the isotropic RAS antenna. The RA.769 threshold is marked by a dashed red line. The crossing point of the curves with the threshold defines the necessary separation distance. We note that for IMT networks of the previous generations, a spurious power level of $-30$~dBm/MHz was in place in Europe. To allow for comparison, the figures also contain the results for this lower level.

\subsection{Setup II: Deployment of 5G devices}\label{subsec:setup2}

More important than the single-interferer analysis is the calculation of the received aggregated power levels, which stem from the full IMT network as the sum of powers from each individual device. Here one aims to use a realistic scenario for the deployment of the IMT equipment to the best of one's knowledge. This not only involves modelling how the BS and UE are distributed in a given area, but one must also account for the quasi-random orientations that the BS and especially the user devices may have. The latter can e.g. be freely rotated with respect to the BS.

Both, BS and UE will use phased-array technology to form beams into the direction of the associated communication partner.  Relatively high gains can be achieved this way, even with the small antenna apertures. BS, which are associated to several user devices simultaneously, use the time-division duplex (TDD) method to serve the UE. In each time slot, the phased-array weights are updated to form a beam to the currently active device. The BS antenna frames themselves are fixed, in contrast to the user devices, where the system needs to form the beam depending on the location relative to the BS and its current rotation state. The beam forming also has an impact on the power that is transmitted into the direction of the RAS station, because the beam direction will significantly change the side-lobe pattern of the antennas (compare Fig.~\ref{fig:compositepatterns}).

Due to the TDD, each user device is only active during 20\% of the time. For BS the TDD factor is set to 80\%. Furthermore, in reality the network as a whole never runs at full capacity. For scenarios such as the one analysed here, \citetalias{itu_tg5_1_doc36_att2} recommends to work with a network loading factor of 20\%.

In \citetalias{itu_tg5_1_doc36_att2} typical number densities are provided, which shall be used for generic compatibility studies involving large areas. The densities differ for BS and UE, but also for the various zone types. The parametrization is done in terms of the percentage of the considered area that has housing, $R_\mathrm{b}=5\%$, and the fraction within this area that is expected to have 5G installations, $R_\mathrm{a}$. For urban zones, $R_\mathrm{a}^\mathrm{urb}=7\%$, while in suburban areas $R_\mathrm{a}^\mathrm{sub}=3\%$. Within the so-defined regions, 30~BS (100~UE) for urban and 10~BS (30~UE) for suburban zones are expected to be installed on average. For suburban open-space zones, the numbers are 1~BS (30~UE).

\begin{figure}[t]
\includegraphics[width=8.3cm]{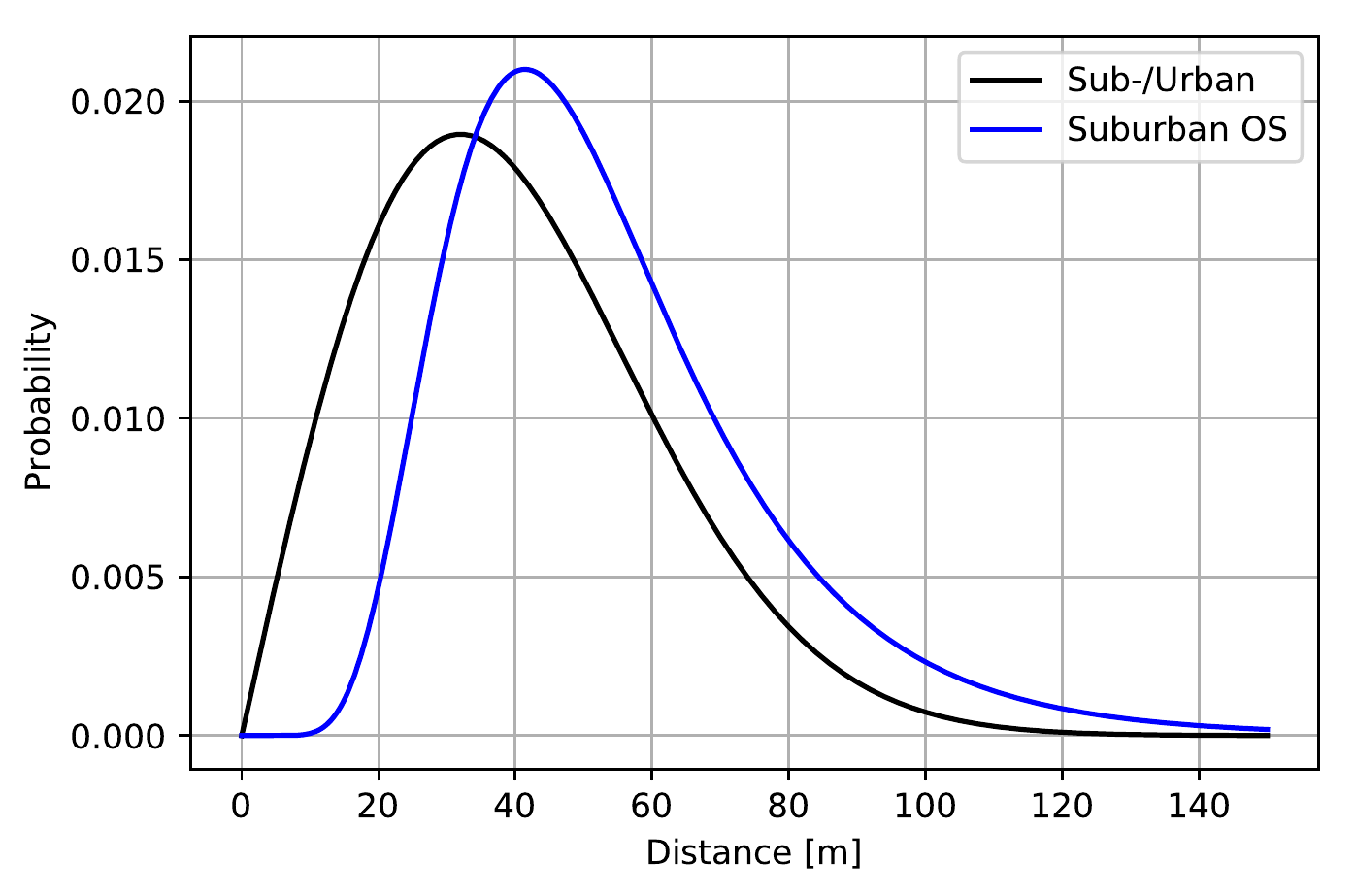}
\caption{Distribution for UE as a function of distance from associated BS.}
\label{fig:ue_bs_dists}
\end{figure}

The final ingredient is the location of the UE relative to the associated BS. \citetalias{itu_tg5_1_doc92_annex01} defines distribution functions for the distance (Rayleigh for urban/suburban; Log-Normal for suburban open-space) and azimuthal position (Normal, $\mathcal{N}(0^\circ, 30^\circ)$) relative to the antenna normal vector\footnote{The azimuthal distribution is clipped for angles larger than $\pm60^\circ$.}. The two distance distributions are displayed in Fig.~\ref{fig:ue_bs_dists}. It should be noted that each BS features only one sector antenna, in accordance to \citetalias{itu_tg5_1_doc36_att2}.

In \citetalias{itu_m2101_0} several possible deployment topologies are discussed, e.g., hexagonal or Manhattan-style grid layouts. In the particular case that is analysed here, the network topology can be neglected because one needs to average over a very large region, such that the aggregated power at the RAS station will be dominated by the (constant) deployment densities defined in \citetalias{itu_tg5_1_doc36_att2}.

\subsection{Aggregated power levels}\label{subsec:aggregation}

We use a Monte-Carlo sampling approach to calculate the aggregated powers from the IMT network. This means that random values for BS and UE positions and rotations are generated from the distribution functions discussed in Section~\ref{subsec:setup2}. This process is repeated many times and for each realization the total power received at the RAS station is calculated. From the ensemble of received powers (one value per trial run), one can then study distribution properties such as the median or average power level, but also the uncertainties (spread about the median).

\begin{figure}[t]
\includegraphics[width=8.3cm,clip=,viewport=354 5 765 330]{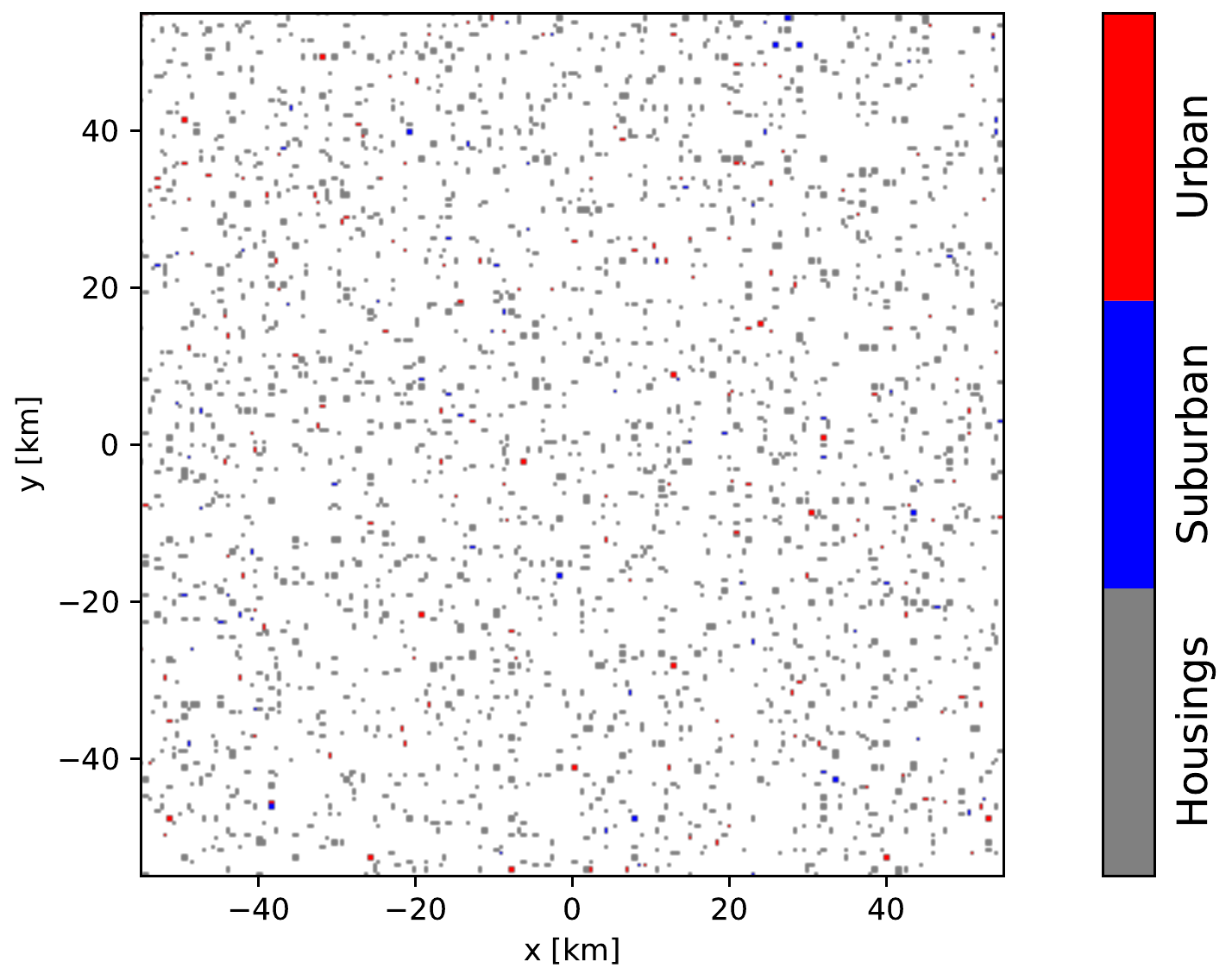}
\caption{Zone types (urban and suburban) randomly assigned to grid cells in the simulated box following a uniform distribution function.}
\label{fig:deployment_uniform}
\end{figure}

The first step in our aggregation analysis is to generate a map containing zone types. A sufficiently large box of size $400~\mathrm{km}\times400~\mathrm{km}$, with grid cells of size $0.5~\mathrm{km}\times0.5~\mathrm{km}$ is created. For each cell, a random number from a uniform distribution, $\mathcal{U}(0,1)$, is drawn. Cells with values between $1-(R_\mathrm{a}^\mathrm{urb}+R_\mathrm{a}^\mathrm{sub})R_\mathrm{b}$ and $1-R_\mathrm{a}^\mathrm{urb}R_\mathrm{b}$ are classified as suburban, while values larger than $1-R_\mathrm{a}^\mathrm{urb}R_\mathrm{b}$ define urban areas. (Regions with housings are defined by values exceeding $1-R_\mathrm{b}$.) Figure~\ref{fig:deployment_uniform} shows a zoom-in to the central part of the simulated box, with the random zone types assigned to it.

\begin{figure*}[t]
\includegraphics[width=16cm]{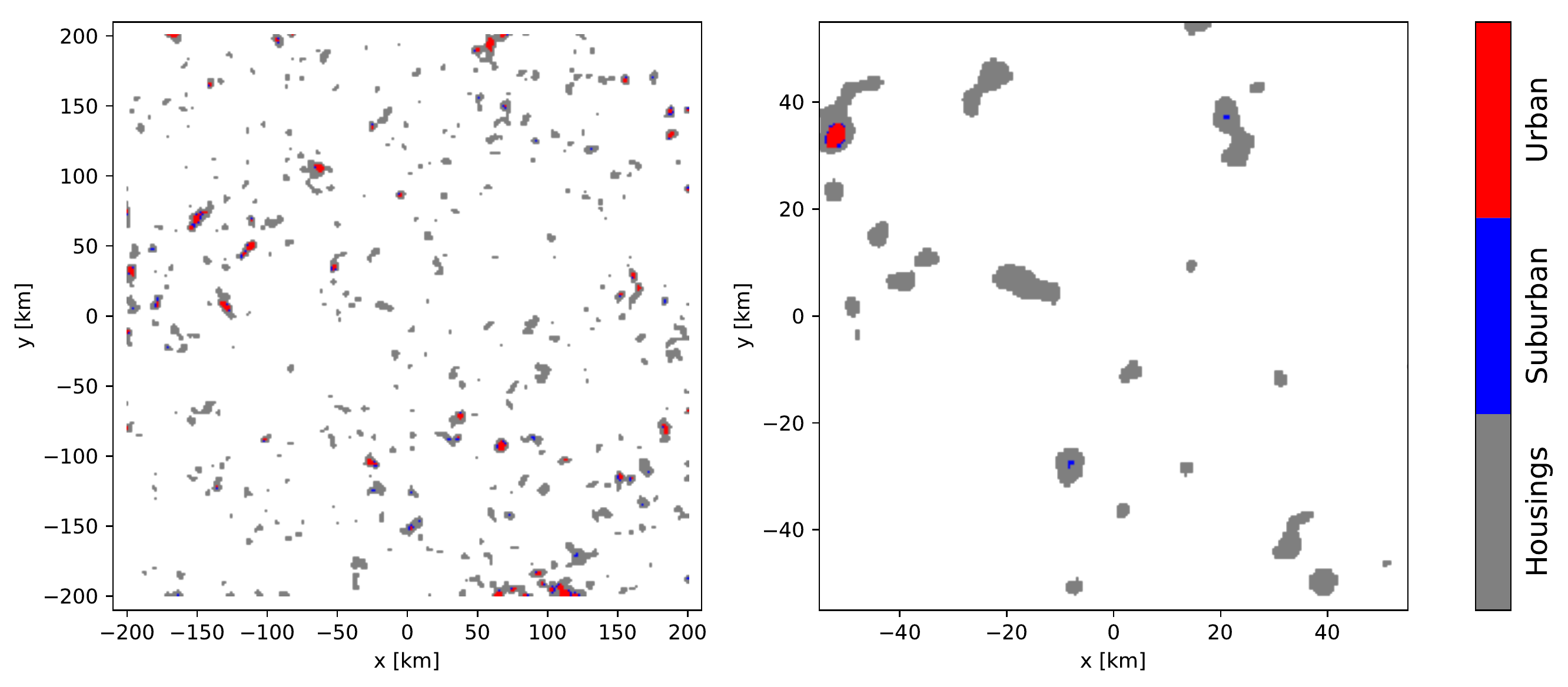}
\caption{Zone types (urban and suburban) randomly assigned to grid cells in the simulated box based on the clustering algorithm described in the text. The right panel shows the full box, the right panel a zoom-in to the central area.}
\label{fig:deployment_clustered}
\end{figure*}

One could argue that such uniform-density deployment is not very realistic. Therefore, as an alternative we develop an algorithm that produces zones, which better resemble a real-world distribution of urban and suburban areas (inspired by the situation around the Effelsberg observatory). For this, some kind of clustering has to be introduced. One possibility to achieve that, is by smoothing the random numbers with one or more Gaussian filter kernels before assigning the zone types. The smoothing introduces a correlation between neighbouring pixels. The larger the smoothing radius, the stronger the clustering of zone types will be. We find that a combination of three different kernels with scales $\sigma_\mathrm{k}=2$, 5, and 15~km and relative amplitudes of 30\%, 30\%, and 40\% can produce a zone-type map, which seems to be quite realistic; see Fig.~\ref{fig:deployment_clustered}.

Into the grid cells, BS are now sampled according to the zone types and given number densities. For each BS, a random azimuthal orientation of the antenna frame is assigned, and the tilting angle in elevation is considered. UE has to be handled differently, because several user devices can be associated to each BS, and the locations of the UE must be in the respective forward cone of the BS. Therefore, for each BS, user devices are sampled into the forward cone, until the total number of UE fits the desired global number density (per zone). UE can be freely rotated, with the only restriction, that the angular distance between antenna normal vector and direction to the associated BS shall not exceed $60^\circ$; see \citetalias{itu_tg5_1_doc92_annex01}. Such a general rotation can for example be constructed from three Euler-angle rotations. Initially, the UE antenna normal shall point to the BS. We first rotate randomly about the UE--BS vector, then about the axis that is perpendicular to the UE--BS vector, is parallel to the $x$--$y$ plane (the ground) and includes the UE position. Finally, another rotation about the UE--BS vector is applied. The first and last rotation angles are drawn uniformly from the interval $\left[-180^\circ,180^\circ\right]$, while the second rotation has random angles from $\left[-60^\circ,60^\circ\right]$. The resulting rotation has the desired properties.

\begin{figure*}[t]
\includegraphics[width=12cm]{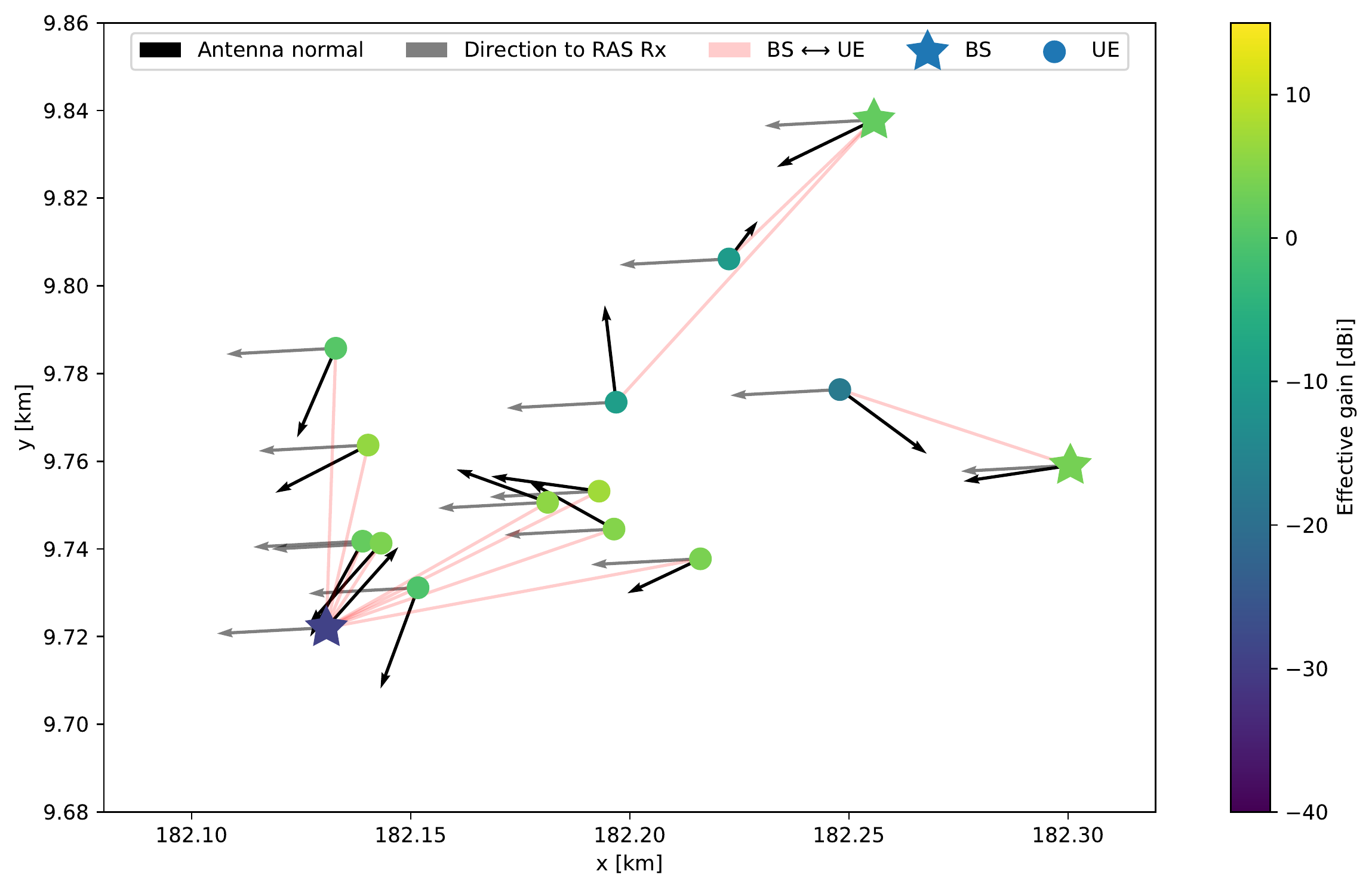}
\caption{Effective antenna gains into the direction of the RAS Rx. The BS and UE are colourized according to the gain. Red solid lines show BS--UE vectors. Black arrows indicate the antenna frame normal vectors, while gray arrows point to the RAS Rx.}
\label{fig:effective_gains}
\end{figure*}

The next step is to calculate the effective antenna gains of each 5G device towards the RAS station. Since the positions and orientations of all devices are known, one can easily determine the azimuth and elevation of the UE--BS vectors, as well as of the direction to the RAS Rx, in the (rotated) antenna frames. Strictly speaking, if the single-element antenna patterns are used, the UE--BS vectors are irrelevant. However, they will be needed for the power control algorithm, which is discussed below. Using the antenna gain formula provided in \citetalias{itu_m2101_0} (implemented in \texttt{pycraf.antenna}), the effective gains follow directly. For BS, there is not a single resulting gain value, but several -- one for each user device in the forward cone. Therefore, the average value is calculated and applied in the following. Figure~\ref{fig:effective_gains} visualizes the geometry and the determined effective gains for a small sub-region that includes three BS and their associated UE. The red lines mark which user device belongs to which BS. The black arrows are the antenna normal vectors (projected into the $x$--$y$ plane, i.e., shorter arrows have larger $z$ components), while grey arrows point to the RAS station. The more the black and grey arrows are aligned, the higher the effective gain. If the composite patterns would be used, the arrows additionally had to be aligned with the red BS--UE vectors to produce highest effective gain. However, the beam forming makes the latter alignment less important. Only if the beam direction has large angular separation from the antenna normal, the achievable gain towards the beam direction drops significantly. However, the specific side-lobe patterns play an important role in the composite-antenna case.

The path propagation loss to the RAS Rx is then computed for each 5G device, as discussed in Section~\ref{subsec:pathloss_5g}, such that all necessary quantities to determine the total power level are now known. There is one last aspect, which makes the calculation slightly more complex: the power control algorithm mentioned in Section~\ref{subsec:setup1_5g}. For this, the link budget between BS and UE has to be determined (according to \citetalias{3gpp_tr_38.901}, UMi -- Street Canyon scenario). The details are beyond the scope of this paper. If the coupling loss between BS and UE is low, a user device may decrease the transmitter power to save battery power (according to a formula provided in \citetalias{itu_m2101_0}). The power control leads to a somewhat lower aggregated UE power, but since BS have a much larger impact on the total power (and are not subject to power control), the change is insignificant.

\begin{figure}[t]
\includegraphics[width=8.3cm]{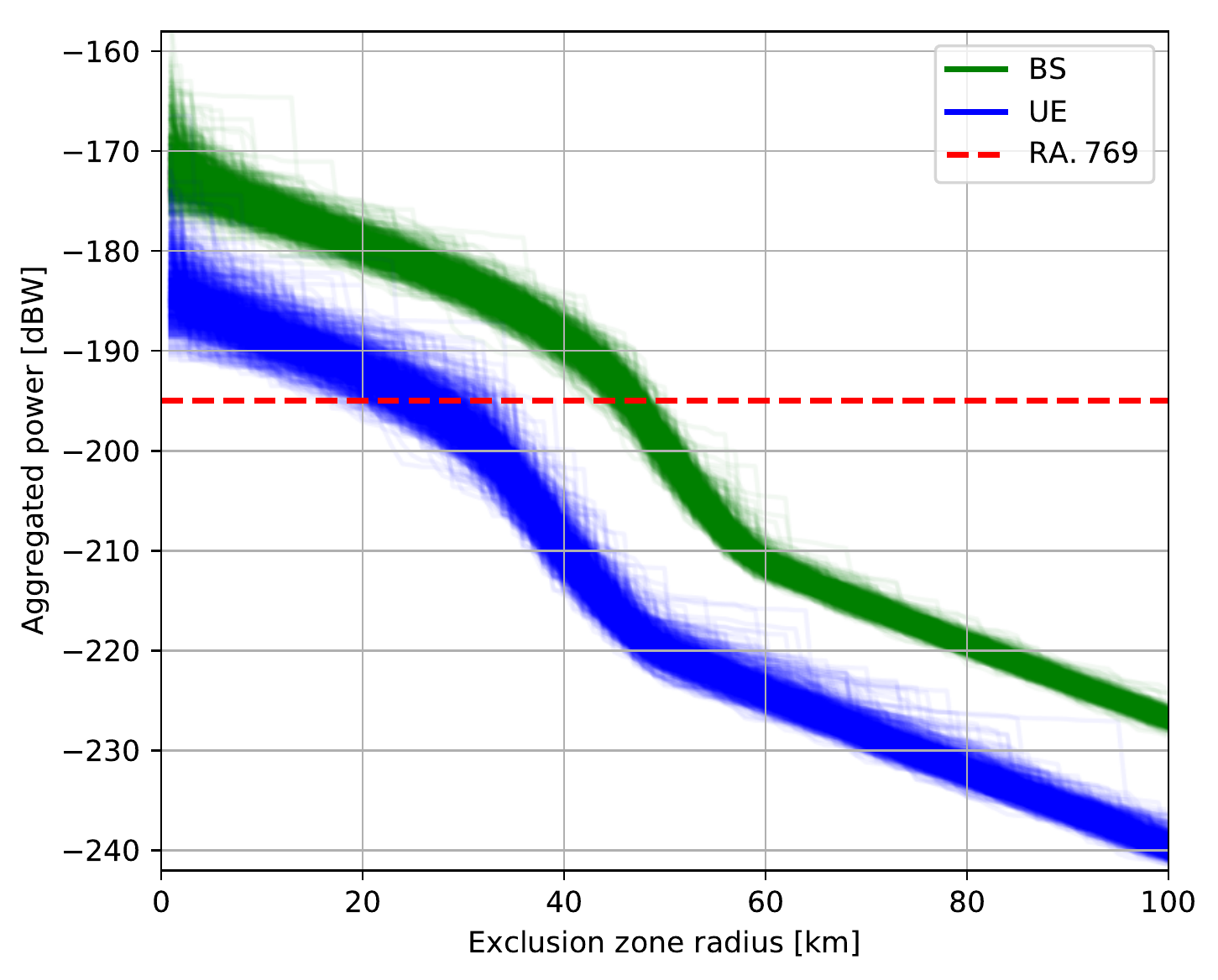}
\caption{Cumulative aggregated power as a function of exclusion zone radius, $r_i$ for uniform distribution of housings. The results of all 1000 Monte-Carlo trials are shown.}
\label{fig:aggpower_uniform_all}
\end{figure}

\begin{figure}[t]
\includegraphics[width=8.3cm]{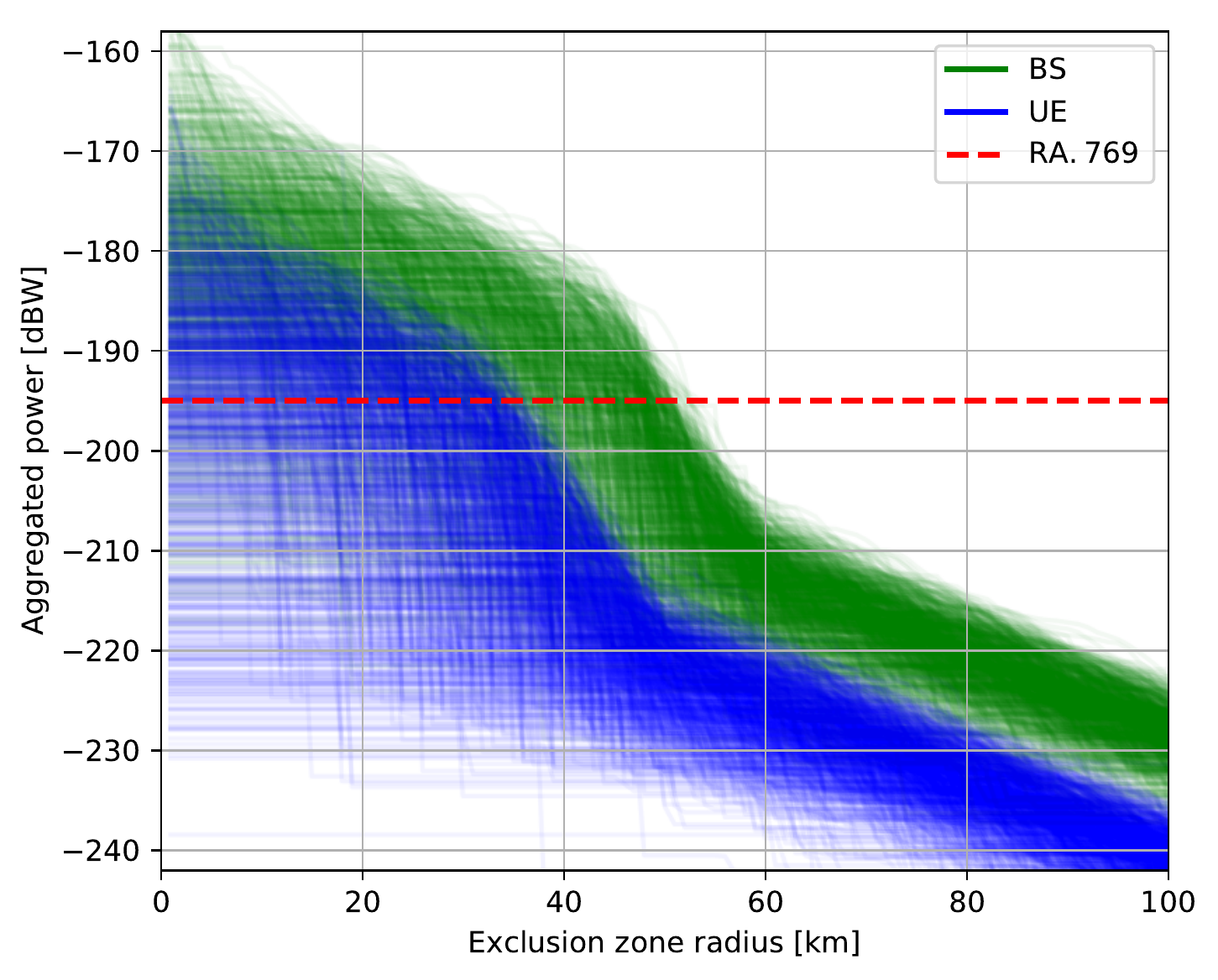}
\caption{As Fig.~\ref{fig:aggpower_uniform_all}, but for the clustered distribution of housings.}
\label{fig:aggpower_clustered_all}
\end{figure}

The aggregated powers received at the RAS station exceed the \citetalias{itu_ra769_2} thresholds in every single Monte-Carlo run. This is not unusual because there is always a high chance that some devices get placed too close to the RAS Rx. To find out, how big an exclusion zone (i.e., the lower limit of separation distances to each device) would need to be, we show cumulative plots of the aggregated powers in Figs.~\ref{fig:aggpower_uniform_all} and \ref{fig:aggpower_clustered_all}. For increasing radii, $r_i$, devices within a sphere with this radius are left out from the aggregation. The larger $r_i$, the lower becomes the received power, as only more distant devices contribute. In total, one thousand runs were carried out. For each of these realizations, Fig.~\ref{fig:aggpower_uniform_all} contains the cumulative distribution for both, BS and UE, in the uniform-density scenario. Figure~\ref{fig:aggpower_clustered_all} shows the results for the clustered-density approach, revealing much higher scatter.

\begin{figure}[t]
\includegraphics[width=8.3cm]{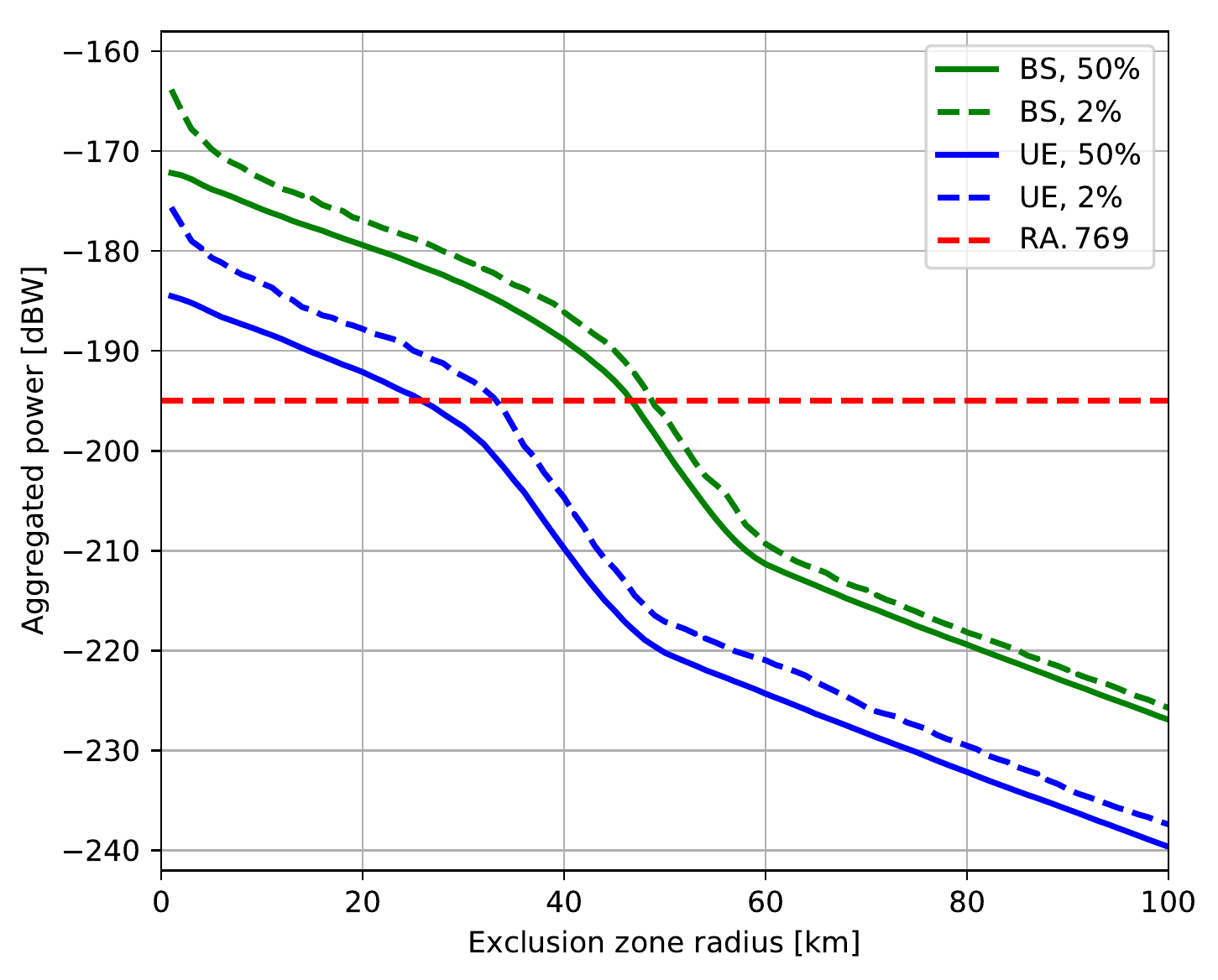}
\caption{Ensemble percentiles (50\% and 98\%) for the distribution of aggregated powers shown in Fig.~\ref{fig:aggpower_uniform_all}.}
\label{fig:aggpower_uniform_percentiles}
\end{figure}

\begin{figure}[t]
\includegraphics[width=8.3cm]{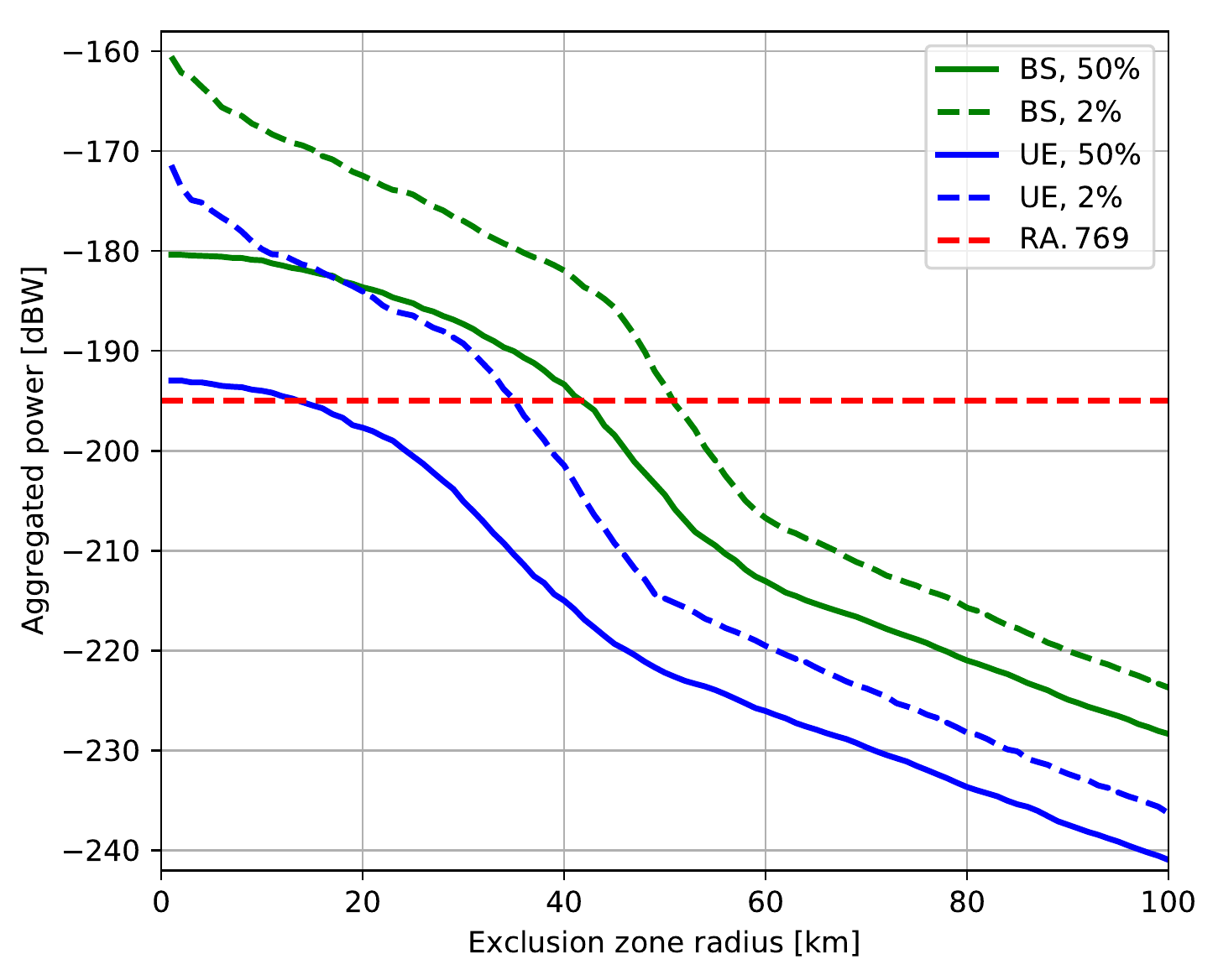}
\caption{As Fig.~\ref{fig:aggpower_uniform_percentiles}, but for the clustered distribution of housings.}
\label{fig:aggpower_clustered_percentiles}
\end{figure}

In order to determine a single value for the necessary exclusion zone radius, we compute distribution percentiles at 50\% (the median) and 98\% (the top 2\% level); see Figs.~\ref{fig:aggpower_uniform_percentiles} and \ref{fig:aggpower_clustered_percentiles}. As can be seen, BS produce on average more than 10~dB higher aggregation levels than UE. Furthermore, there is a steep regime between 30~km and 60~km, where the power level drops quickly. This behaviour can be attributed to the path propagation loss, which shows a similar kink (compare Figs.~\ref{fig:pathloss_bs} and \ref{fig:pathloss_ue}). Hence, the transmitted power levels could be increased substantially, without the exclusion zone size having to grow quickly (unless one would not want to accept separation distances of $\sim$60~km in the first place).

\conclusions\label{sec:conclusions}  
The \texttt{pycraf} package proves to be a valuable and versatile tool to assist spectrum managers with accomplishing compatibility studies. In simpler cases, this only needs a few lines of Python code, but being a programming library, \texttt{pycraf} is also useful for very complex analyses, e.g., full Monte Carlo simulations of mobile communication networks.

Such a case, a compatibility study involving next-generation 5G at 24~GHz and a RAS telescope, was presented in detail. In the generic case (flat-earth, i.e., zero terrain heights) the aggregated power levels produced at the radio-astronomical receiver would exceed the maximally permitted thresholds defined in \citetalias{itu_ra769_2} unless an exclusion zone with a radius of few tens of kilometres would be established.


\codeavailability{The \texttt{pycraf} library is made available on the Python package distribution server PyPI (\textit{Python Package Index}, \url{https://pypi.python.org/pypi/pycraf}). It is open source software licensed (GNU General Public License v3.0). The source code is hosted on the GitHub platform (\url{https://github.com/bwinkel/pycraf}) and is also available on Zenodo (\url{https://doi.org/10.5281/zenodo.1244192}). Contributions are welcome.} 













\competinginterests{The authors declare that they have no conflict of interest.} 


\begin{acknowledgements}
We thank Uwe Bach for carefully proof-reading our manuscript and providing valuable comments.
Furthermore, we would like to express our gratitude to the developers of the many C/C++ and Python libraries, which are made available as open-source software and which we used: most importantly, NumPy \citep{NumPy}, SciPy \citep{SciPy}, Cython \citep{Cython}, and Astropy \citep{Astropy}. Figures were prepared using matplotlib \citep{Matplotlib}.
\end{acknowledgements}







\bibliographystyle{copernicus}
\bibliography{references.bib}

\end{document}